\documentclass[%
 reprint,
twocolumn,
showpacs,
 amsmath,amssymb,
 aps,
prb,
]{revtex4-1}

\usepackage{graphicx}
\usepackage{dcolumn}
\usepackage{bm}

\newcommand{\ex}{\mathrm{e}}
\newcommand{\im}{\mathrm{i}}

\usepackage{caption}
\begin{document}
%
%
\title{Orbital Magnetism of Graphene Nanostructures: \\
        Bulk and Confinement Effects} 
%
\author{Lisa He\ss{}e}
\author{Klaus Richter}
\affiliation{Institut f\"ur Theoretische Physik, Universit\"at Regensburg, D-93040 Regensburg, Germany
}%
%
%
%

\date{\today}

\begin{abstract}
We consider the orbital magnetic properties of non-interacting charge carriers in graphene-based 
nanostructures in the low-energy regime. The magnetic response of such systems results both, from
bulk contributions and from confinement effects that can be particularly strong in ballistic quantum dots. 
First we provide a comprehensive study of the magnetic susceptibility $\chi$ of bulk graphene 
in a magnetic field for the different regimes arising from the relative magnitudes of the energy scales 
involved, {\em i.e.} temperature, Landau level spacing and chemical potential.  
We show that for finite temperature or chemical potential, $\chi$ is not divergent although the 
diamagnetic contribution $\chi_{0}$ from the filled valance band exhibits the well-known $-B^{-1/2}$ 
dependence.  We further derive oscillatory modulations of $\chi$, corresponding to de Haas-van 
Alphen oscillations of conventional two-dimensional electron gases. 
These oscillations can be large in graphene, thereby compensating the diamagnetic contribution 
$\chi_{0}$ and yielding a net paramagnetic susceptibility for certain energy and magnetic field regimes. 
Second, we predict and analyze corresponding strong, confinement-induced susceptibility oscillations in
graphene-based quantum dots with amplitudes distinctly exceeding the corresponding bulk susceptibility.
Within a semiclassical approach we derive generic expressions for orbital magnetism of
graphene quantum dots with regular classical dynamics. Graphene-specific features can be traced 
back to pseudospin interference along the underlying periodic orbits.
We demonstrate the quality of the semiclassical approximation by comparison with quantum
mechanical results for two exemplary mesoscopic systems, a graphene disk with infinite mass-type 
edges and a rectangular graphene structure with armchair and zigzag edges, using
numerical tight-binding calculations in the latter case.

\end{abstract}

\maketitle



\section{\label{sec:Introduction}Introduction}

Since the seminal work of Landau~\cite{springerlink:10.1007/BF01397213} it is known that a conventional free electron gas exihbits
a weak diamagnetic orbital magnetic response. In two dimensions (2d) and at low magnetic field, its
magnetic susceptibility $\chi$ is just a constant, i.e.\ independent of Fermi energy and $B$-field. 
For Dirac fermions in 2d, e.g.\ charge cariers in graphene close to the charge neutrality point, 
the situation is different: As McClure showed nearly 50 years ago~\cite{PhysRev.104.666}, a non-interacting 2d 
system of massless Dirac fermions features a Curie-type $1/k_{\rm B}T$ behavior\cite{PhysRevB.75.115123} at finite temperature $T$ 
that merges, for vanishing temperature, into a peculiar dependence on the chemical
potential\cite{PhysRev.104.666, PhysRevB.20.4889, PhysRevB.75.235333, PhysRevB.76.113301,  PhysRevLett.102.177203, JPSJ.80.114705, 1751-8121-44-27-275001, PhysRevB.83.235409, PhysRevB.80.075418}
: $\chi \sim \delta(\mu)$, i.e.\ a magnetic response that is divergent in the undoped limit
and otherwise zero.

In this work we pose the question how orbital magnetism in graphene-based nano- and mesoscale systems 
is altered through the presence of the confinement. Similar questions had been intensively discussed in the early 
nintees for small disordered metallic rings \cite{Webb}, quasi ballistic micron-sized rings \cite{Mailly} 
and square cavities \cite{Levy} based on conventional 2d electron systems. The magnetic response of 
(ensembles of) these mesoscopic systems, namely the observed persistent current in the rings and 
the susceptibility of the cavities, turned out to exceed the bulk Landau 
diamagnetism by one to two orders of magnitude. These original experimental findings triggered broad 
theoretical activities (for reviews see \cite{Schwab,Richter,Ullmo3}) investigating in particular also 
the role of non-interacting versus interacting contributions to the orbital magnetism. 
While twenty years ago further progress in the field had been hindered by experimental limitations, 
recent new high-precision cantilever magnetization (persistent current) measurements of ensembles of 
rings proved \cite{Harris} the feasibility to 
reliably measure orbital magnetism of nanoscale objects. The results of these recent experiments 
are essentially in line with earlier theory based on non-interacting systems \cite{Oppen}.
Given the peculiar orbital magnetic behavior of bulk graphene, and in view of the above mentioned possibility to 
observe confinement-enhanced magnetism in nanostructures~\cite{Harris}, it hence is of interest to explore 
also orbital magnetism in graphene nanostructures, a topic that has been barely addressed in the literature. 

Here, we employ a trajectory-based semiclassical path integral formalism to compute the orbital 
magnetic susceptibility. As recently shown, such an approach is suitable for the quantitative 
description and interpretation of the density of states \cite{PhysRevB.84.075468} and conductance \cite{PhysRevB.84.205421} of 
graphene-based cavities. This approach allows for the incorporation of graphene-specific boundary effects 
(zigzag, armchair and infinite mass). The confinement geometry and the type of edge is then encoded 
in the amplitudes and phases of paths (hitting the boundaries) that enter into  the respective 
semiclassical trace formulae. We combine this approach with an earlier semiclassical treatment of 
orbital magnetism in conventional ballistic electron cavities \cite{PhysRevLett.74.383,Richter}. 
We show that the susceptibility of graphene cavities of linear system 
size $R$ exhibits confinement-induced oscillations in $k_{\rm F} R$ where $k_{\rm F}$
 is the Fermi momentum. For integrable geometries and at low temperatures their amplitude is  parametrically 
larger by a factor of $\sqrt{k_{\rm F}R}$ than the corresponding bulk susceptibility. However, 
graphene cavities additionally carry features of bulk graphene. Hence, in the first part of the paper 
we include a comprehensive discussion of graphene bulk orbital magnetism. While a number of previous 
works addressed various parameter regimes separately we aim at a systematic presentation of the various bulk 
regimes.

\begin{figure}[htbp]
 \centering
\includegraphics[width = 0.45\textwidth]{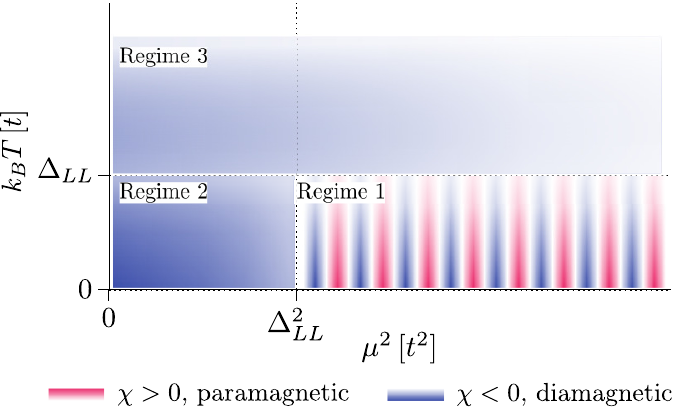}
\caption[]{Rough schematic overview of the orbital magnetic behaviour of graphene in a 
perpendicular field for the energy regimes studied in \mbox{Subsec.~\ref{sec:Main_1_therm}}. Here
$\mu$ is the chemical potential, $k_{B} T$ the thermal energy and 
$\Delta_{LL}\propto \sqrt{B}$ represents the Landau level spacing.  
Blue (red) regions refer to parameter regimes where graphene shows diamagnetism (paramganetism).  
The colour intensity roughly indicates the strength of the magnetic response.  
The magnetic susceptibility $\chi$ is diamagnetic in allmost all areas except the de Haas-van Alphen 
regime, $\mu > \Delta_{LL} > k_{B}T$, with susceptibility oscillations 
of $\chi$ being linear in $\mu^2$ (and $1/B$).
}
\label{fig:overview}
\end{figure}

This is simplistically sketched in Fig.~\ref{fig:overview}. It shows an overall diamagnetic behavior
up to the energy region governed by de Haas-van Alphen oscillations\cite{onsager, Peierls, PhysRevB.60.R11277, PhysRevB.69.075104, 0953-8984-22-11-115302} for $k_{\rm B}T < \Delta_{\rm LL} <
\mu$, with $\Delta_{\rm LL}$ proportional to the Landau level spacing. 
However, the diamagnetic regions exhibit interesting parametrical dependences that we will derive and review. 
For instance, the afore mentioned divergent behavior of $\chi$ at $T\!=\!0$ is smoothed out if $k_{\rm B}T$
is bigger than the mean level spacing.

The paper is organized as follows: 
After summarizing the necessary thermodynamic formalism in Sec.~\ref{sec:Basic}, 
we first give a comprehensive account of bulk magentism in graphene in Sec.~\ref{sec:Main_1},
addressing the various parameter regimes mentioned above. This also involves introducing our numerical approach 
and our scheme to extract bulk results from the numerics performed for finite systems.
In the other main Sec.~\ref{sec:Main_2} we consider in detail finite-size effects in the orbital
magnetic response of nanostructued graphene. There we generalize the existing semiclassical approaches
to quantitatively describe and interprete oscillatory effects in the susceptibility. These semiclassical
predictions are compared to corresponding quantum calculations for disc-like and rectangluar geometries.  
We focus on integrable structures since chaotic or diffusive geometries are expected to exhibit a parametically 
weaker magnetic response.  


\section{\label{sec:Basic}Basic thermodynamic quantities}

In order to investigate the orbital magnetic properties of a quasi-two dimensional solid  
in general, it is convenient to start from the total grand potential in the presence of a perpendicular
magnetic field of strength $B$,
\begin{align}
\label{eq:i1}
 \hspace{-0.3cm}\Omega(\mu, B) = -\frac{1}{\beta} \! \int\limits_{-\infty}^{\infty} \! \mathrm{d}E \,
		 \rho(E, B) \ln \left[ 1 + \ex^{-\beta \left(E-\mu\right) } \right], 
\end{align}
where \mbox{$1/\beta = k_{\mathrm{B}} T$} denotes the thermal energy.
The chemical potential $\mu$ is assumed to be $B$-independent.
The total density of states \cite{enderlein1997fundamentals} (DOS), 
$\rho(E, B)=\rho_{v}(E, B) + \rho_{c}(E, B)$, comprises 
conduction and valence band states simultaneously
as well as the field-dependence of the energy spectrum of the solid. 
Defining $E_{v / c}$ as the energy of the band edge of the valence/conduction band, the corresponding densities 
of states fullfil
\mbox{$\rho_{v / c}(E, B) = 0,$ $\forall E \gtrless E_{v / c}$},
even for a vanishing energy gap \mbox{$E_{g} = E_{c} - E_{v} = 0$} as in the 
case of graphene.
Without loss of generality, 
$\mu$ is chosen to be larger than $E_v$.
Due to the properties of the total DOS, the grand potential can be decomposed as 
$\Omega = \Omega_{v} + \Omega_{c}$, where
\begin{align}
\label{eq:i2}
 \Omega_{v}(\mu, B) = - \frac{1}{\beta} \int\limits_{-\infty}^{E_{v}} \! \mathrm{d}E\,
		 \rho_{v}(E, B) \ln \left[ 1 + \ex^{-\beta \left(E-\mu\right) } \right],
\end{align}
\begin{align}
\label{eq:i3}
 \Omega_{c}(\mu, B) =  - \frac{1}{\beta}\int\limits_{E_{c}}^{\infty} \! \mathrm{d}E \,
		 \rho_{c}(E, B) \ln \left[1 + \ex^{-\beta \left(E-\mu\right) } \right].
\end{align}
Equation (\ref{eq:i3}) contains the contribution to $\Omega$ from electrons in the conduction band for 
Fermi energies $\mu > E_{v}$ or thermal excitation.  In the limit $T\rightarrow 0$ only states with energy 
$E_{c} \leq E \leq \mu$ are occupied.  
In view of
\begin{align}
\label{eq:i4}
 -\lim\limits_{\beta \rightarrow \infty}\frac{1}{\beta} \ln\left(1 + \ex^{-\beta x}\right) = x \, \theta\left(-x\right),
\end{align} 
and taking the limit $T \! \rightarrow \! 0$ in \mbox{Eq.~(\ref{eq:i2})}, yields 
the contribution to $\Omega$ from the completely filled valence band:
\begin{align}
 \label{eq:i5}
 \Omega_{0}(\mu, B) =  \int\limits_{-\infty}^ {E_{v}}  \! \mathrm{d}E \, \rho_{v}(E, B) (E - \mu).
\end{align}  
In general, the integral (\ref{eq:i5}) can diverge,  if the particular model assumes a valence band without lower boundary.  
As we will discuss in \mbox{Subsec.~\ref{sec:Main_1_cond}} for bulk graphene in the low energy approximation, $\Omega_{0}$ 
can be decomposed into a $B$-field-dependent and a divergent part, which does not include any field-dependence and therefore 
has no effect on the magnetic properties.  

By pulling a
factor  $\exp{\left[-\beta(E-\mu)\right]}$ out of the logarithm in \mbox{Eq.~(\ref{eq:i2})} $\Omega_{v}$ can be represented as 
\begin{align}
 \hspace{-0.5cm}\label{eq:i6}
\begin{split}
\Omega_{v}(\mu, B) =\ & \Omega_{0}(\mu, B)\\
		&- \frac{1}{\beta} \! \int\limits_{-\infty}^{E_{v}} \! \mathrm{d}E \, 
		\rho_{v}(E, B) \ln \left[ 1 + \ex^{\beta \left(E-\mu\right) } \right]\!.
\end{split}\hspace{-0.5cm}
\end{align}
The second term in \mbox{Eq.~(\ref{eq:i6})} contains a similar contribution 
to $\Omega$ as $\Omega_{c}$ corresponding to electron vacancies 
at finite temperature.
As a first conclusion, $\Omega$ can be decomposed into the $T$-independent part $\Omega_{0}$, coming from the filled part of the valence band, 
and a contribution
\begin{align}
\hspace{-0.5cm}\label{eq:i7}
  \Omega_{T}(\mu, B) =&\ \Omega(\mu, B) - \Omega_{0}(\mu, B)\\
\label{eq:i8}
\begin{split}
		=& - \frac{1}{\beta} \! \int\limits_{-\infty}^{\infty} \! \! \mathrm{d}E \, 
		\left\{
		\rho_{v}(E, B) \ln \left[ 1 + \ex^{\beta \left(E-\mu\right) } \right]
		\right.\\
		&\left. \qquad \qquad + 
		\rho_{c}(E, B) \ln \left[ 1 + \ex^{-\beta \left(E-\mu\right) } \right]
		\right\}
\end{split}\hspace{-0.5cm}
\end{align}
due to excited electrons in the conduction band and holes in the valence band.  
Within the relevant temperature range the integral (\ref{eq:i8}) converges fast due to the
exponential decay of the integrand at both integration limits.

The total magnetic susceptibility is defined as
\begin{align}
 \label{eq:i9}
 \chi(\mu, B) =- \frac{\mu_{0}}{\mathcal{A}} \left(\frac{\partial^{2}\Omega(\mu, B)}{\partial B^2} \right)_{T,
\mu} \, .
\end{align}
In view of \mbox{Eq.~(\ref{eq:i7})}, it can be decomposed into 
\begin{align}
 \label{eq:i10}
 \chi(\mu, B) =\ \chi_{0}(\mu, B) + \chi_{T}(\mu, B),
\end{align} 
with
\begin{align}
 \label{eq:i11}
 \hspace{-0.5cm}\chi_{\mathrm{x}}(\mu, B) = - \frac{\mu_{0}}{\mathcal{A}} \left(\frac{\partial^{2}
		\Omega_{\mathrm{x}}(\mu, B)}{\partial B^2} \right)_{T, \mu}\!\!\!\!,\ \ 
		\mathrm{x} = 0,\, T.
\end{align}
Here, $\mathcal{A}$ denotes the area of the system and $\mu_0$
is the vacuum permeability.  
As will be shown in \mbox{Subsec.~\ref{sec:Main_1_cond}} for bulk graphene, 
$\chi_{0}$, 
which is of similar origin as the Landau susceptibility \cite{springerlink:10.1007/BF01397213, Richter} 
of non-relativistic electron gases, represents a smooth, diamagnetic contribution $\propto 1/\sqrt{B}\ $
\cite{PhysRevB.69.075104, PhysRevB.75.115123, PhysRevB.80.075418, PhysRevB.75.235333}
to the total susceptibility.
Contrarily, $\chi_{T}$ in Eq.~(\ref{eq:i10}) 
can yield an oscillatory contribution to $\chi$ for certain energy regimes.  
In bulk systems this oscillatory behavior refers to the 
de Haas-van Alphen effect \cite{springerlink:10.1007/BF01338364, onsager}, 
whereas in finite systems additional modulations in $\chi$ occur as signatures of the confinement, see
\mbox{Sec.\ \ref{sec:Main_2}}.


\section{\label{sec:Main_1}Bulk orbital susceptibility}

\subsection{\label{sec:Main_1_dens}Spectral properties of Landau quantized charge carriers with linear dispersion}
In this section the orbital magnetic properties of bulk graphene in the energy range of linear dispersion are discussed.  
The graphene sheet is assumed to lie in the $x$-$y$-plane perpendicular to an external, homogeneous
$B$-field.
Then the energies of the charge carriers
are Landau quantized \cite{springerlink:10.1007/BF01397213}.  
The  Landau levels of massless Dirac-Weyl particles in 2D describing bulk graphene read
\cite{JPSJ.74.777, PhysRevB.76.081406, 10.1038/nphys653} 
\begin{align}
 \label{eq:landau}
E_{n} = \mathrm{sgn}(n)\frac{ \sqrt{2}\hbar v_{F}}{l_{B}}\sqrt{\left|n\right|} \, ,
\end{align}
with \mbox{$n \in \mathbb{Z}$}.  
Here, \mbox{$l_{B} = \sqrt{\phi_{0}/(2\pi B)}$} denotes the magnetic length with the magnetic flux quantum \mbox{$\phi_{0} = h/e$}.  
Every Landau level $E_{n}$ has a twofold spin degeneracy $g_{s}$ and valley degeneracy $g_{v}$ as well as a 
\mbox{$\varphi = \phi/\phi_{0}$}-fold degeneracy \mbox{($\phi = B\,\mathcal{A}$)} which can be, e.g., deduced from phase space arguments \cite{onsager, kittel} and 
Bohr-Sommerfeld quantization \cite{messiah1991quantenmechanik} 
of the corresponding cyclotron orbits.  Thus the orbital degeneracy in graphene is identical to that of 
Landau levels of ordinary 2D electron gases\cite{springerlink:10.1007/BF01397213}, \mbox{$\epsilon_{n} = (\hbar/m^{*})\,eB\,(n+1/2)$}, with effective mass $m^{*}$ and \mbox{$n \in \mathbb{N}_{0}$}.    
In this case the lowest Landau level has the finite value \mbox{$\epsilon_{0} = (\hbar/m^{*})\,eB/2$} while 
for graphene 
\mbox{$E_{0}=0$} attains zero and lies precisely at the touching point of conduction and valence band.  
In the presence of a magnetic field conduction and valence band states occupy the zeroth Landau level equally leading to an increase of the total energy 
of the filled valence band.  Thus the contribution $\chi_0$ from the filled valence band is expected to be 
diamagnetic as discussed in detail in \mbox{Subsec.~\ref{sec:Main_1_cond}}.  
Whether the total susceptibility $\chi$, \mbox{Eqs.~(\ref{eq:i9}, \ref{eq:i10})}, 
is para- or diamagnetic depends on the contribution $\chi_{T}$ of excited electrons and holes 
in the particular energy regime.

The single-particle DOS of bulk graphene,
\begin{align}
\label{eq:m1_12}
 \rho(E, B) = g\,\varphi \sum\limits_{n=-\infty}^{\infty} \delta\left(E - E_{n}(B)\right),
\end{align}
can be decomposed into a smooth and an oscillatory part with respect to $E$ and $B$.  
By means of Poisson summation\cite{brack_sc} of the Landau index $n$ one obtains
\begin{align}
\label{eq:m1_13}
 \hspace{-0.5cm}\rho(E, B) &= C \left|E\right| \left[
		1 + 2 \sum\limits_{m=1}^{\infty}\cos\left(\pi m \left(\frac{E\, l_{B}}{\hbar v_{F}}\right)^{2}\right)\right]\\
\label{eq:m1_14}
	    &= \bar{\rho}(E) + \rho^{\mathrm{osc}}(E, B)
\end{align}
with \mbox{$C = g \mathcal{A}/[2\pi (\hbar v_{F})^{2}]$} and \mbox{$g = g_{s} g_{v}$}.  
Note that each term in \mbox{Eqs.~(\ref{eq:m1_13}, \ref{eq:m1_14})} and thereby the total DOS reflects particle-hole symmetry, i.e.~\mbox{$\rho(E, B) = \rho(-E, B)$}, 
due to the nearest neighbor hopping approximation underlying the effective Dirac hamiltonian.  
The smooth part \mbox{$\bar{\rho}(E) = C \left|E\right|$} is $B$-independent and identical to the bulk DOS of the field free system \cite{RevModPhys.81.109}.  
Hence the entire contribution to $\chi$ arises from the oscillatory part $\rho^{\mathrm{osc}}(E, B)$ that
can be rewritten as
\begin{align}
 \label{eq:m1_15}
\begin{split}
 \hspace{-0.5cm} \rho^{\mathrm{osc}}(E, B) = g\,\varphi \sum_{n = 1}^{\infty}\left[\delta\left(E - E_{n}(B)
\right) + \delta\left(E + E_{n}(B)\right)\right]\\
    \hspace{-1cm} + g\,\varphi\,\delta\left(E\right) - C \left|E\right|.
\end{split}
\end{align}
This represention clearly indicates that the orbital magnetism arises only from Landau levels 
with $n \neq 0$.  The zeroth Landau level leads to a $\varphi$-linear contribution to $\Omega$ and thus does not
contribute to $\chi$.  As for the DOS 
the related thermodynamic potentials can be decomposed into
\begin{align}
\label{eq:m1_16}
\Omega(\mu, B) 
		  =  \bar{\Omega}(\mu) + \tilde{\Omega}(\mu, B).
\end{align}
Each term in \mbox{Eq.~(\ref{eq:m1_16})} can be further split as shown in \mbox{Eqs.~(\ref{eq:i7}, \ref{eq:i10})}, i.e. 
\mbox{$X = X_{0} + X_{T}$}, where \mbox{$X = \bar{\Omega}, \tilde{\Omega}$}.
Note that $\bar{\Omega}$ arises directly from the field independent bulk DOS $\bar{\rho}(E)$, and hence
\mbox{$\chi \propto \partial^{2}\Omega/\left(\partial B^{2}\right) = \partial^{2}\tilde{\Omega}/\left(\partial B^{2}\right)$}.  
We will show below that though $\tilde{\Omega}$ arises from the oscillatory part of the DOS, it yields not only an oscillatory but 
also a smooth contribution to the susceptibility.


\subsection{\label{sec:Main_1_comp}Comparability of numerical results with analytical bulk DOS calculations}

The comparison between the analytical results for bulk graphene, to be discussed in \mbox{Subsec.~\ref{sec:Main_1_cond}} 
and \ref{sec:Main_1_therm}, with the numerical tight-binding data of confined graphene quantum dots will
demonstrate the importance of bulk effects in finite structures. 
Moreover, vice versa, we will employ the numerical calculations, restricted to finite gemetries, to
confirm the results from the effective bulk theory based on the Dirac equation. For such a comparison
we need to extract the bulk contribution from the numerical results in an appropriate way as discussed
below.

\begin{figure}[h]
\includegraphics[width = 0.45\textwidth]{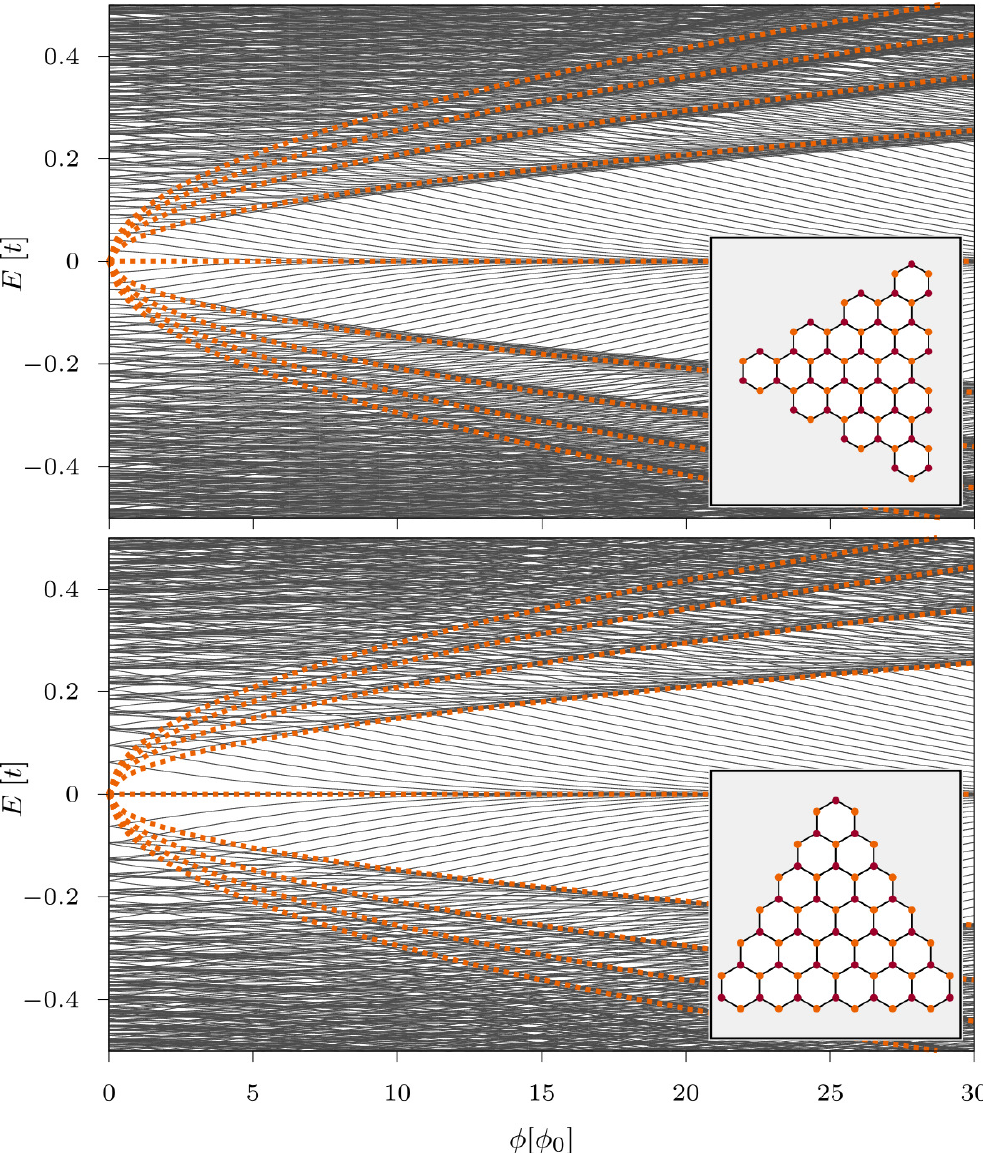}
\caption{(Color online) Energy spectrum of triangular quantum dots with armchair edge (upper panel) and
zigzag edge (lower panel) as a function of the normalized magnetic flux through 
         each system. The dashed green lines refer to the lowest Landau energies $E_{n}$.  
	    In the case of zigzag geometry the zigzag edge states clearly appear close to $E \approx 0$ and
contribute to the zeroth Landau level. Insets: Sketch of the geometries, the actual systems considered are
much larger: ${\cal L} \simeq 100 a$.
}
\label{fig:spectrum_qd}
\end{figure}

The finite systems considered have an equilateral triangular geometry with either pure armchair or zigzag boundaries.  
This particular choice of geometry enables also a distinct analysis of edge effects due to zigzag
boundaries.
Each system has mesoscopic dimensions, i.e. the triangle side lengths are \mbox{$\mathcal{L} \approx 100\,a$}, where $a$ is the graphene lattice constant, such that the region of linear dispersion contains enough energy 
levels to require good comparability with the theory.  
The eigenenergies of the triangles are calculated within tight-binding
approximation\cite{epub12142,Wimmer} including only nearest neighbor hopping $t$ and using 
the Lanczos algorithm \cite{arpack1, ARPACK}.  Figure \ref{fig:spectrum_qd} shows the resulting energy 
spectrum for conduction and valence band energies \mbox{$\left|E\right| \leq 0.55\,t$} as a 
function of the normalized magnetic flux \mbox{$\phi/\phi_{0}$}. One can clearly see the 
condensation of the eigenenergies into Landau levels\cite{PhysRevB.81.245411, PhysRevB.84.245403}for fluxes $\phi > 5\,\phi_{0}$.  
This is the regime where bulk effects should be distinctly observable in the finite systems.  

In \mbox{Subsec.~\ref{sec:Main_1_cond}} we calculate the contribution $\chi_{0}$ from the valence band in Dirac approximation, which corresponds to the 
Landau susceptibility\cite{} of electron gases.  
Therefore we assume an unbounded valence band with linear dispersion which does not reflect the real band
structure of graphene further away from the Dirac point.  
For this reason, $\chi_{0}$ is not accessible within tight-binding approximation even if one would go to very large system sizes.  
Thus the comparison of the analytic theory with numerical data for the quantum dots is restricted to the temperature dependent 
part of  $\tilde{\Omega}$ and $\chi$, respectively, where only the energy levels close to the Fermi level
contribute.

\begin{figure}[htbp]
\centering
\includegraphics[width = 0.5\textwidth]{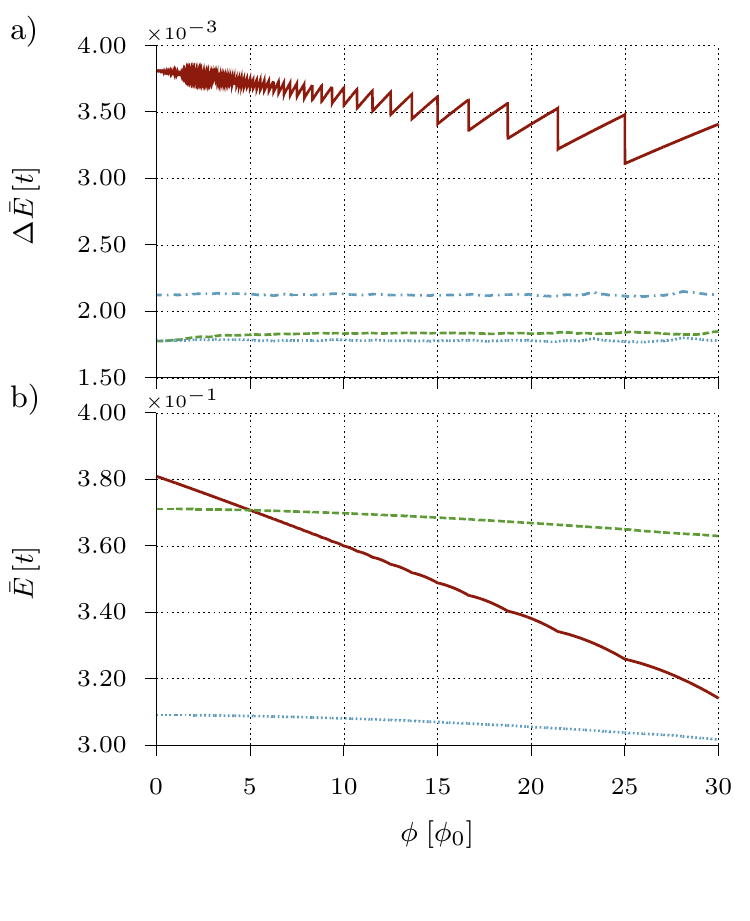}
\caption[Mean level spacing] {(Color online) a) Comparison of the mean level spacing (in units of hopping energy
$t$) of the lowest $600$ electron states of the two triangluar quantum dots,
see \mbox{Fig.~\ref{fig:spectrum_qd}}, 
with the Dirac model for bulk graphene (red solid line) calculated from \mbox{Eq.~(\ref{eq:landau})} 
as a function of the magnetic flux.  
The blue dotted (dashed-dotted) line represents the mean level spacing of the zigzag system with (without)
considering edge states.
Due to the edge states  $\Delta E$ is smaller than in the armchair system (green dashed line).
b) Mean energy of the lowest $600$ electronic energies as a function of the magnetic flux. 
The full (red), dotted (blue) and dahed (green) lines correspond to the bulk, zigzag and armchair system,
respectively.
}
\label{fig:mean_level_spacing}
\end{figure}

As one can deduce from \mbox{Fig.~\ref{fig:spectrum_qd}} the mean level spacings \mbox{$\Delta\bar{E}^{(\mathrm{ac}, \mathrm{zz})}$} 
of the finite systems differ from the mean Landau level spacing \mbox{$\Delta \bar{E}^{(\mathrm{bulk})}$}
of the bulk system.
In Fig.~\ref{fig:mean_level_spacing}a) we compare explicitly \mbox{$\Delta\bar{E}^{(\mathrm{x})}$} 
(\mbox{x $=$ ac, zz, bulk}), calculated from the first $600$ electronic states for each case, as a function of the normalized magnetic flux.  
When dealing with thermodynamic potentials and related observables, finite temperature $T$, encoded in the Fermi-Dirac statistics, implies an  
effective broadening $1/\beta$ of each energy level as can be seen from Eq.~(\ref{eq:i8}). For an
appropriate comparison of the Dirac-type bulk theory with the tight-binding results 
for the finite-size structures the thermal energies chosen should obey
\begin{align}
 \label{eq:m1_17}
 \hspace{-0.1cm}\beta^{(\mathrm{bulk})}\Delta\bar{E}^{(\mathrm{bulk})}(\phi) \approx \beta^{(\mathrm{x})}\Delta\bar{E}^{(\mathrm{x})}(\phi), \ \  \mathrm{x} = \mathrm{ac}, \mathrm{zz}.
\end{align}
To get reliable values from this expression the edge states are not considered in the case of the zigzag system
since they lead to underestimating the mean level spacing.

To compare the properties of the bulk system with those of the quantum dots for finite magnetic flux it
is necessary to average Eq.~(\ref{eq:m1_17}) over the flux interval considered.  
The resulting level spacings averaged over $\phi \in \left[0, 30\,\phi_{0}\right]$ can be read off 
from \mbox{Table \ref{tab:table1}}. 
The above procedure, providing an adequate comparison between all three systems, refers to 
the entire thermodynamic potentials and related properties. Independently, 
the individual energy levels for each of the considered systems can be written as $E^{(\mathrm{x})}_{i} = \bar{E}^{(\mathrm{x})} + \delta E^{(\mathrm{x})}_{i}$, where 
\mbox{$\bar{E}^{(\mathrm{x})} = 1/N\,\sum_{i = 1}^{N} E^{(\mathrm{x})}_{i}$} denotes the mean energy of the $N$ valence band states considered.  
\mbox{Figure \ref{fig:mean_level_spacing}b)} shows the flux dependence of $\bar{E}^{(\mathrm{x})}$ for
each system averaged over the lowest $N=600$ electron states.
In all three cases the mean energies are of the same order of magnitude. 
Due to the contribution of edge states, 
\mbox{$\bar{E}^{\mathrm{ac}} > \bar{E}^{\mathrm{zz}}$}.

The grand potential for each system reads
\begin{align}
 \label{eq:m1_18}
 \hspace{-0.2cm}\Omega^{(\mathrm{x})}(\mu, B) = - \frac{1}{\beta} \sum\limits_{i} \ln\!\left[1 + \ex^{-\beta\left(\bar{E}^{(\mathrm{x})} + \delta E^{(\mathrm{x})}_{i} - \mu\right)}\right]\!\!.
\end{align}
The properties of the exponential function and the logarithm yield a rough scaling behavior of \mbox{$\Omega^{(\mathrm{ac}, \mathrm{zz})}/\Omega^{(\mathrm{bulk})} \approx \gamma^{(\mathrm{ac}, \mathrm{zz})}$} 
for each system reflecting in first approximation
\begin{align}
 \label{eq:m1_19}
 \beta^{\mathrm{ac, zz}}\bar{E}^{(\mathrm{ac, zz})} -\beta^{\mathrm{bulk}}\bar{E}^{(\mathrm{bulk})} \gtrless 0.
\end{align}
Resulting differences in the absolute value of $\Omega$ and $\chi$,
respectively, for fixed $\mu$ and $\varphi$ can be approximately compensated by rescaling the bulk value 
with the factor $\gamma^{(\mathrm{ac}, \mathrm{zz})}$.  
The factors \mbox{$\gamma^{(\mathrm{ac}, \mathrm{zz})}$} were obtained by fitting using the Levenberg-Marquardt \cite{springerlink:10.1007/BFb0067700} algorithm.  
 
As a consequence of \mbox{$\bar{E}^{\mathrm{ac}} > \bar{E}^{\mathrm{zz}}$},
\mbox{Fig.~\ref{fig:mean_level_spacing}(b)}, and \mbox{Eq.~(\ref{eq:m1_18})} 
we expect the susceptibility contribution $\chi_{T}$ for a zigzag triangular quantum dot to be smaller than for the corresponding armchair system at same temperature 
corresponding to \mbox{Eq.~(\ref{eq:m1_19})}.  
This behavior is also confirmed in \mbox{Ref.~[\onlinecite{PhysRevB.87.115433}]}, where the orbital magnetic
properties of hexagonal and triangular graphene nanostructures are numerically studied within tight-binding approximation.  

\renewcommand{\arraystretch}{1.3}
\begin{table}[t]
\begin{ruledtabular}
\begin{tabular}{c|ccc}
\textrm{x}&
\textrm{bulk}&
\textrm{armchair}&
\textrm{zigzag}\\
\colrule
$\langle\Delta \bar{E}^{(\mathrm{x})}\rangle_{\phi}\ \left[10^{-3}\,t\right]$ & 3.514 & 1.828 &  2.124\ (1.780)\\
\colrule
\textrm{$\langle\bar{E}^{(\mathrm{x})}\rangle_{\phi}\ \left[t\right]$} &  0.349 & 0.368 & 0.306
\end{tabular}
\end{ruledtabular}
\caption{\label{tab:table1}
Flux average of the mean energy \mbox{$\langle\bar{E}^{(\mathrm{x})}\rangle_{\phi}$} and mean level spacing \mbox{$\langle\Delta \bar{E}^{(\mathrm{x})}\rangle_{\phi}$} for the first $600$ electron states of bulk graphene, 
an armchair and a zigzag triangular quantum dot, same as Fig.~\ref{fig:spectrum_qd}. The considered flux interval
amounts to  \mbox{$\left[0, 30\,\phi_{0}\right]$}. The number in parenthesis
 comprises the edge states.
}
\end{table}

\subsection{\label{sec:Main_1_cond}Susceptibility contribution from filled valence band}

As discussed in \mbox{Subsec.~\ref{sec:Main_1_dens}}, the susceptibility contribution $\chi_{T}$  from the filled
valence band, \mbox{Eq.~(\ref{eq:i5})}, can be evaluated from \mbox{Eq.~(\ref{eq:i8})} by using only the 
field-dependent part of the DOS, $\rho^{\mathrm{osc}}(E, B)$:
\begin{align} 
 \hspace{-0.5cm}\label{eq:m1_20} 
  \tilde{\Omega}_{0}(\mu, B) =& \int\limits_{-\infty}^{0}\mathrm{d}E \, \rho^{\mathrm{osc}}(E, B) (E-\mu)\\
\label{eq:m1_21}
\begin{split}
			       =& -2C\sum\limits_{m=1}^{\infty}\mathrm{Re}\Bigg[
				   \lim\limits_{\eta \rightarrow 0}
				   \int\limits_{0}^{\infty}\mathrm{d}E \left(E^{2} + \mu E\right)\\
				     &\times\ex^{-\left[\eta - \im \pi m \left(\frac{l_{B}}{\hbar v_{F}}\right)^{2}\right]\,E^{2}}\Bigg].
\end{split}\hspace{-0.5cm}
\end{align}
Solving this integral and taking the limit \mbox{$\eta \rightarrow 0$} yields
\begin{align}
 \label{eq:m1_22}
 \tilde{\Omega}_{0}(B) = \frac{K}{2}\varphi^{3/2} \sum\limits_{m=1}^{\infty}\frac{1}{m^{3/2}} 
			       = \frac{K}{2}\varphi^{3/2} \zeta\left(\frac{3}{2}\right)\!,
\end{align}
where all prefactors are absorbed in the constant 
\begin{align}
  \label{eq:m1_23}
 K = 4 \sqrt{\pi}\, C \left(\frac{\hbar v_{F}}{\sqrt{\mathcal{A}}}\right)^{3}
                      = 2 g\frac{\hbar v_{F}}{\sqrt{\mathcal{A} \pi}}.
\end{align}  
Indeed, $\Omega_{0}$ and thereby the corresponding susceptibility 
\begin{align}
 \label{eq:m1_24}
 \hspace{-0.5cm}
  \chi_{0}(B) = - \frac{\mu_{0}g}{\phi_{0}^{2}}\hbar v_{F} \sqrt{\frac{\mathcal{A}}{\pi}}\frac{3\zeta\left(\frac{3}{2}\right)}{4} \frac{1}{\sqrt{\varphi}} 
        \propto -\frac{1}{\sqrt{B}}
\end{align}
are independent of the chemical potential. $\chi_0(B)$ is diamagnetic because the grand potential 
of the valence band, \mbox{$\bar{\Omega}_{0} + \tilde{\Omega}_{0}(B)$}, increases in the presence 
of a perpendicular magnetic field, i.e.\ \mbox{$\tilde{\Omega}_{0}(B) > \tilde{\Omega}_{0}(0)$}.
The susceptibility $\chi_{0}$ diverges as $1/\sqrt{B}$ implying that small variations of the flux cause huge
changes in the magnetization of bulk graphene in the low-field regime.  
The scaling behavior (\ref{eq:m1_24}) of $\chi_{0}$ was first 
discovered by McClure in 1956 within his studies of the diamagnetic properties of graphite\cite{PhysRev.104.666} and confirmed 
by various research groups\cite{PhysRevB.69.075104, PhysRevB.75.115123, PhysRevB.80.075418, PhysRevB.75.235333, PhysRevB.86.125440} for monolayer graphene.  
In \mbox{Sec. \ref{sec:Main_1_therm}} we show that 
this singularity of $\chi_{0}$, however, need not lead to a divergence of the total 
susceptibility \mbox{$\chi = \chi_{0} + \chi_{T}$}.\\  
In the case of a bulk 2DEG the quantity corresponding to $\chi_{0}$ 
is the Landau susceptibility \cite{Richter}
\begin{align}
 \label{eq:m1_25}
  \chi_{L} = - \mu_{0} g_{s} \frac{\pi}{6}\frac{\hbar^{2}}{\phi_{0}^{2} m^{*}} \, .
\end{align}
It is also independent of $\mu$ but moreover does not depend on $B$.  
To estimate the relative strength of graphene diamagnetism we consider the 
ratio $\chi_{0}/\chi_{L}$ which reads
\begin{align}
 \label{eq:m1_26}
  \frac{\chi_{0}(B)}{\chi_{L}} \approx 0.2\; \frac{m^{*}}{m_{\mathrm{e}^{-}}} \sqrt{ \frac{\mathcal{A}}{\varphi}}\;\mathrm{nm}^{-1}.
\end{align}
To give an explicit example, consider GaAs \mbox{($m^{*} = 0.067\,m_{\mathrm{e}^{-}}$)} and a graphene flake with a typical length \mbox{$\mathcal{L}/l_{B} \gg 1$}, such that 
bulk effects dominate over finite-size signatures in $\chi$.  
Choosing typical values \mbox{$\phi = \phi_{0}$} and \mbox{$\mathcal{A} \approx 100^{2}\;\mathrm{n m}^{2}$}, 
Eq.~(\ref{eq:m1_26}) yields \mbox{$\chi_{0} \approx \chi_{L}$}, i.e.\ the diamagnetic contribution from the valence band in graphene is 
comparable to the Landau susceptibility of a 2DEG for a magnetic field of $B \approx 0.5\,\mathrm{T}$.

\subsection{\label{sec:Main_1_therm}Susceptibility contribution from thermally excited charge carriers}

To investigate the contribution to $\chi$ from excited electrons and holes
we start from \mbox{Eq.~(\ref{eq:i8})} considering only the field-dependent 
part $\rho^{\mathrm{osc}}(E, B)$ of the DOS in \mbox{Eq.~(\ref{eq:m1_13})}:
\begin{eqnarray}
 \label{eq:m1_27} 
 &&\tilde{\Omega}_{T}(\mu, B) = -\frac{1}{\beta}\int\limits_{-\infty}^{\infty}\mathrm{d}E \, \rho^{\mathrm{osc}}(E, B)\\
		&&\times \left\{\theta(-E)\ln\left[1+\ex^{\beta\left(E-\mu\right)}\right] 
		+ \theta(E)\ln\left[1+\ex^{-\beta\left(E-\mu\right)}\right]\right\}.\nonumber
\end{eqnarray} 
Due to the integration over energy and the temperature dependence of $\tilde{\Omega}_{T}$ the corresponding susceptibility $\chi_{T}$ can contain a smooth as well as an oscillatory 
part, which is directly accessible within the semiclassical description of finite-size contributions to $\chi$
as shown in Sec.\ \ref{sec:Main_2}.  

For the following considerations it is useful to integrate Eq.~(\ref{eq:m1_27}) twice by parts yielding
\begin{align}
 \label{eq:m1_28}
 \tilde{\Omega}_{T}(\mu, B) =  \int\limits_{-\infty}^{\infty}\mathrm{d}E \, \mathcal{N}(E, B) f'\left(E-\mu\right),
\end{align}
with the integral over particle number fluctuations,
\begin{align}
 \label{eq:m1_29}
 \mathcal{N}(E, B) =& \int\limits_{0}^{E}\mathrm{d}E'\int\limits_{0}^{E'}\mathrm{d}E'' \rho^{\mathrm{osc}}(E, B)\\
\label{eq:m1_30}
                   =& K\varphi^{3/2}\sum\limits_{m=1}^{\infty} 
		      \frac{\mathrm{S}\left(\sqrt{\pi m} \frac{\left|E\right|\,l_{B}}{\hbar
v_{F}}\right)}{m^{3/2}} \, .
\end{align}
Here, \mbox{$\mathrm{S}(x) =  \sqrt{2/\pi}\int_{0}^{x}\mathrm{d}t\,\sin\left(t^{2}\right)$}
is  the Fresnel integral\cite{gradshtein} and
$K$ is defined by \mbox{Eq.~(\ref{eq:m1_23})}.
In \mbox{Eq.~(\ref{eq:m1_28})} 
\begin{align}
 \label{eq:m1_31}
 f'\left(x\right) = 
		-\frac{\beta}{4} \mathrm{sech}^{2}\left(\frac{\beta}{2}x\right)  \xrightarrow{\beta \rightarrow \infty} -\delta(x)
\end{align}
denotes the derivative of the Fermi distribution function \mbox{$f\left(x\right) = \left[1+\exp\left(\beta\,x\right)\right]^{-1}$}.  
We rewrite  \mbox{Eq.~(\ref{eq:m1_28})} as
\begin{align}
 \label{eq:m1_32}
   \tilde{\Omega}_{T}(\mu, B) =   K\varphi^{3/2}\sum\limits_{m = 1}^{\infty} \frac{\omega_{m, T}(\mu, B)}{m^{3/2}},
\end{align}
where $\omega_{m, T}$ is defined as the energy integral
\begin{align}
  \label{eq:m1_33}
  \hspace{-0.5cm}\omega_{m, T}(\mu, B) = \!\!\!\int\limits_{-\infty}^{\infty}\!\!\!\mathrm{d}E\,\mathrm{S}\!\left(\!\sqrt{\pi m}\frac{\left|E\right|\,l_{B}}{\hbar v_{F}}\!\right)f'\left(E - \mu\right)\!.
\end{align}
Since the integral (\ref{eq:m1_33}) cannot generally be solved analytically, it is convenient to discuss separately different regimes defined through the ratios between 
the relevant length scales entering the problem, namely the magnetic length $l_{B}$, the Fermi wavelength
\mbox{$\lambda_{F} = \hbar v_{F}/\mu$}, and the thermal wavelength $\lambda_{T} = \hbar v_{F}\,\beta$.  
For the sake of simplicity we define the two dimensionless parameters 
\begin{align}
 \label{eq:m1_34}
 \alpha = \frac{\lambda_{F}}{l_{B}} \propto \frac{\Delta_{LL}}{\mu},\quad \gamma= \frac{\lambda_{T}}{l_{B}} \propto \frac{\Delta_{LL}}{k_{B}T},
\end{align}
where \mbox{$\Delta_{LL} \propto \sqrt{B}$} denotes the energy spacing between adjacent Landau levels.

\subsubsection{\label{sec:Main_1_therm_1}Regime: $\gamma > 1 > \alpha$}

In this parameter range the Landau level spacing is larger than or comparable to the thermal energy, but smaller than the chemical potential.  The resulting temperature dependent 
contribution to $\Omega$ is therefore expected to show an oscillatory modulation as a function of $\mu$ or
$\varphi$ 
known as de Haas-van Alphen effect in electron gases \cite{onsager, Peierls, PhysRevB.60.R11277}.  
Moreover we will show that the $1/\sqrt{B}$ singularity in Eq. (\ref{eq:m1_24}) is cancelled.
Hence  it is useful to decompose the Fresnel integral in Eq.~(\ref{eq:m1_33}) into its smooth and oscillatory part, i.e. 
\mbox{$\mathrm{sgn}(x)\,\mathrm{S}(x) = 1/2 + \tilde{\mathrm{S}}(x)$}.  
The function $\tilde{\mathrm{S}}(x)$ oscillates around zero and can be written in terms of the hypergeometric function \mbox{$\mathrm{U}(1/2; 1/2, -\im x^{2})$} or its 
integral representation as shown in \mbox{App.~\ref{sec:App_Int_Fresnel}},
\begin{align}
  \label{eq:m1_36}  
  \tilde{\mathrm{S}}(x)=& -\frac{1}{\sqrt{2\pi} }\mathrm{Im} \left[\int\limits_{0}^{\infty}\!\mathrm{d}u \frac{\ex^{-\left(u - \im\right)x^{2}}}{\sqrt{\pi\,u}\left(u - \im\right)}\right].
\end{align}
Then the energy integral (\ref{eq:m1_33}) reads
\begin{align}
  \label{eq:m1_37} 
  \omega_{m, T} = -\frac{1}{2} + \!\!\int\limits_{-\infty}^{\infty}\!\!\mathrm{d}E\,\tilde{\mathrm{S}}\!\left(\!\sqrt{\pi m}\frac{\left|E\right|\,l_{B}}{\hbar v_{F}}\!\right)f'\left(E - \mu\right).
\end{align}
Note that the remaining integral directly leads to the $B$ field-dependent part of the total grand potential, \mbox{$\tilde{\Omega} = \tilde{\Omega}_{0} + \tilde{\Omega}_{T}$}, 
since the first term in \mbox{Eq.~(\ref{eq:m1_37})} exactly cancels with $\tilde{\Omega}_{0}$,
Eq.~(\ref{eq:m1_27}), after inserting it into \mbox{Eq.~(\ref{eq:m1_32})}.  
Then $\tilde{\Omega}$ can be cast into the form
\begin{align}
  \label{eq:m1_38} 
  \hspace{-0.125cm}\tilde{\Omega}(\mu, B) =\sum\limits_{m=1}^{\infty}\!\frac{K\varphi^{3/2}}{\sqrt{2\pi\,m^{3}}} 
      \mathrm{Im}\!\left[\int\limits_{0}^{\infty}\mathrm{d}u\frac{\mathrm{Y}_{T}(\mu, B, u)}{\sqrt{\pi\,u}(u - \im)}\right]\!\!,
\end{align}
with 
\begin{align}
 \label{eq:m1_39} 
 \mathrm{Y}_{T}(\mu, B, u) =\!\!\!\int\limits_{-\infty}^{\infty}\!\!\mathrm{d}E\,f'(E - \mu) \, 
\ex^{-\left(u - \im \right)\pi m\left(\frac{E\,l_{B}}{\hbar v_{F}}\right)^{2}} .
\end{align}
As shown in App.~A of \mbox{Ref.~[\onlinecite{Richter}]} for a similar situation, 
\begin{align}
 \label{eq:m1_41}
 \mathrm{Y}_{T}(\mu, B, u) \approx \mathrm{Y}_{0}(\mu, B, u)\,\mathrm{R}_{T}\left[\mathrm{\phi}'(\mu, B, u)\right],
\end{align}
where the temperature damping factor $\mathrm{R}_{T}$ is defined as
\begin{align}
  \label{eq:m1_42}
\mathrm{R}_{T}\left[\mathrm{\phi}'(\mu, B, u)\right] = \frac{\frac{\pi}{\beta}\mathrm{\phi}'(\mu, B, u)}{
	  \sinh\left[\frac{\pi}{\beta}\mathrm{\phi}'(\mu, B, u)\right]} \xrightarrow{\beta \rightarrow \infty} 1
\end{align}
and results from the derivative of the Fermi-Dirac distribution and \mbox{$\mathrm{\phi}'(\mu, B, u) = \left.\partial\mathrm{\phi}(E, B, u)/(\partial E)\right|_{E = \mu}$}.  
From Eqs.~(\ref{eq:m1_41}, \ref{eq:m1_42}) follows 
\begin{align}
 \label{eq:m1_43}
 \mathrm{Y}_{T}(\mu, B, u) \approx \ex^{-\left(u - \im\right)\pi m/\alpha^{2}} \mathrm{R}_{T}\left(\beta\frac{2\pi m}{\alpha \gamma}\right),
\end{align}
so the field-dependent part of $\Omega$ finally reads 
\begin{align}
  \label{eq:m1_44}
\tilde{\Omega}(\mu, B) 
	&\approx&   \!\!\!\!K\sum\limits_{m=1}^{\infty}\frac{\varphi^{3/2}}{m^{3/2}} \tilde{\mathrm{S}}\left(\frac{\sqrt{\pi\,m}}{\alpha}\right) \mathrm{R}_{T}\left(\beta\frac{2\pi m}{\alpha \gamma}\right)\!.
\end{align}
Compared to the rapid magneto oscillations of $\tilde{\mathrm{S}}$, the factor $\mathrm{R}_{T}$ 
only slowly varies on the relevant scales so that 
its magnetic field derivatives can be neglected in the calculation of the total magnetic susceptibility:
\begin{align}
\label{eq:m1_45}
 \chi(\mu, B) =& -\frac{\mu_{0} g}{\phi_{0}^{2}} \hbar v_{F}\frac{3\sqrt{\mathcal{A}}}{2\pi} \sum\limits_{m = 1}^{\infty}
		\frac{\mathrm{R}_{T}\left(\beta\frac{2\pi m}{\alpha\gamma}\right)}{m^{3/2}}\frac{\mathrm{J}\left(\frac{\sqrt{\pi m}}{\alpha}\right)}{\sqrt{\varphi}}\\
\label{eq:m1_46}
    =&\chi_{0}(B)\times\frac{2}{\sqrt{\pi}\zeta\left(\frac{3}{2}\right)}\sum\limits_{m = 1}^{\infty}
		\frac{\mathrm{R}_{T}\left(\beta\frac{2\pi m}{\alpha\gamma}\right)}{m^{3/2}}\mathrm{J}\left(\frac{\sqrt{\pi m}}{\alpha}\right),
\end{align}
with $\chi_{0}(B)$ defined in \mbox{Eq.~(\ref{eq:m1_24})}.
At finite temperatures the sum in  \mbox{Eq.~(\ref{eq:m1_46})} is exponentially damped due to  $\mathrm{R}_{T}$, 
ensuring convergence of the corresponding expression.  
The function $\mathrm{J}(x)$ is defined as
\begin{align}
\label{eq:m1_47}
 \mathrm{J}(x) = \tilde{\mathrm{S}}\left(x\right) + \sqrt{\frac{2}{\pi}} x \left[\sin\left(x^2\right)- \frac{2 x^2}{3}\cos\left(x^2\right)\right],
\end{align}
yielding $\mu^{2}$- as well as $1/\phi$-periodic oscillations of $\chi$, respectively \mbox{$\chi_{T} = \chi - \chi_{0}$}, which can be extracted from 
\mbox{Eq.~(\ref{eq:m1_46})} and \mbox{Eq.~(\ref{eq:m1_24})}.  
This becomes more obvious by transforming the expression (\ref{eq:m1_47}) for $\mathrm{J}(x)$  into
\begin{align}
 \label{eq:m1_48}
\mathrm{J}(x) \!=\! -\!\frac{\cos\!\left(x^{2}\right)}{\sqrt{2\pi}}\left[\mathrm{\Sigma}_{1}\!\left(x^{2}\right)
\! +\! \frac{4}{3} x^{3}\right]
\!-\!\frac{\sin\!\left(x^{2}\right)}{\sqrt{2\pi}}\!\left[\mathrm{\Sigma}_{2}\left(x^{2}\right) \!-\! 2 x\right].
\end{align}
by defining \mbox{$\mathrm{\Sigma}_{1/2}(x^{2}) \!=\! \mathrm{Im}/\mathrm{Re}[\exp(\im
\pi/4)\,\mathrm{U}(1/2; 1/2; -\im x^{2})]$} and rewriting $\tilde{\mathrm{S}}(x)$,
\mbox{Eq.~(\ref{eq:m1_36})}.
For the magnetization of bulk graphene an expression similar to \mbox{Eq.~(\ref{eq:m1_46})} is derived in
\mbox{Ref.~[\onlinecite{PhysRevB.69.075104}]} considering additionally a band gap and impurity scattering, 
whereas in \mbox{Ref.~[\onlinecite{0953-8984-22-11-115302}]} the effect of an additional in-plane electric 
field is studied.  

\begin{figure}[htbp]
 \centering
\includegraphics[width = 0.5\textwidth]{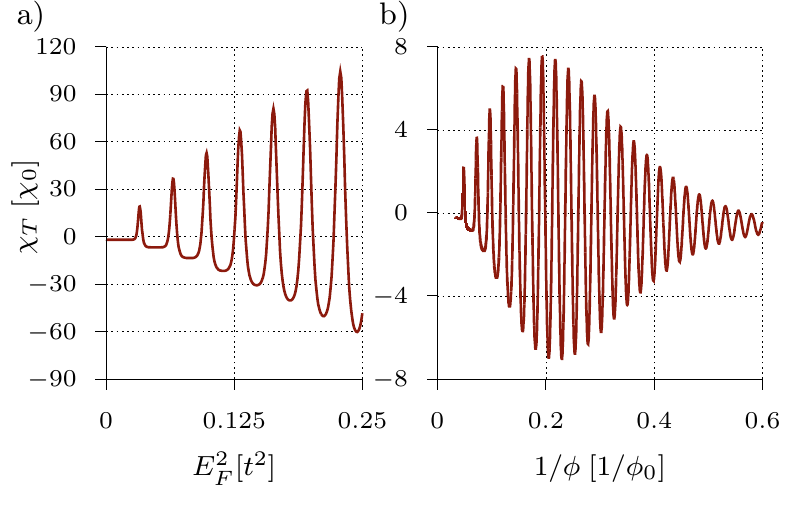}
\caption[]{Susceptibility contribution $\chi_{T}$ for bulk graphene calculated from \mbox{Eqs.~(\ref{eq:m1_45}, \ref{eq:m1_46})}  at \mbox{$1/\beta^{\mathrm{(bulk)}} \approx 3\cdot 10^{-3}\,t$}.  
	  a) $\chi_{T}$ shown as a function of $\mu^{2}$ for fixed \mbox{$\phi = 15\,\phi_{0}$}.
	  b) $\chi_{T}$ plotted as a function of the inverse magnetic
flux for fixed \mbox{$\mu = 0.3\,t$}.
}
\label{fig:new_norm_regime_b_g_1_g_a_III}
\end{figure}

In \mbox{Fig.~\ref{fig:new_norm_regime_b_g_1_g_a_III}} the oscillatory behavior of $\chi_{T}$ is demonstrated.  
In panel a) $\chi_{T}$ exhibits equidistant extrema when plottet as a function of $\mu^{2}$ at \mbox{$\phi = 15\,\phi_{0}$}.  
Panel b) shows the $1/\phi$-periodicity of $\chi_{T}$ at \mbox{$\mu = 0.3\,t$}.  
In both cases the thermal energy is chosen such that \mbox{$1/\beta^{\mathrm{(bulk)}} \approx 3\cdot 10^{-3}\,t$}.  
The amplitude of the $\chi_{T}$ oscillations is about one order of magnitude larger than $|\chi_{0}|$,
implying that the full orbital susceptibility $\chi$ 
of graphene oscillates between strong diamagnetic but also paramagnetic behavior as a function of $\mu$ and $B$, respectively.

\begin{figure}[htbp]
 \centering
\includegraphics[width = 0.5\textwidth]{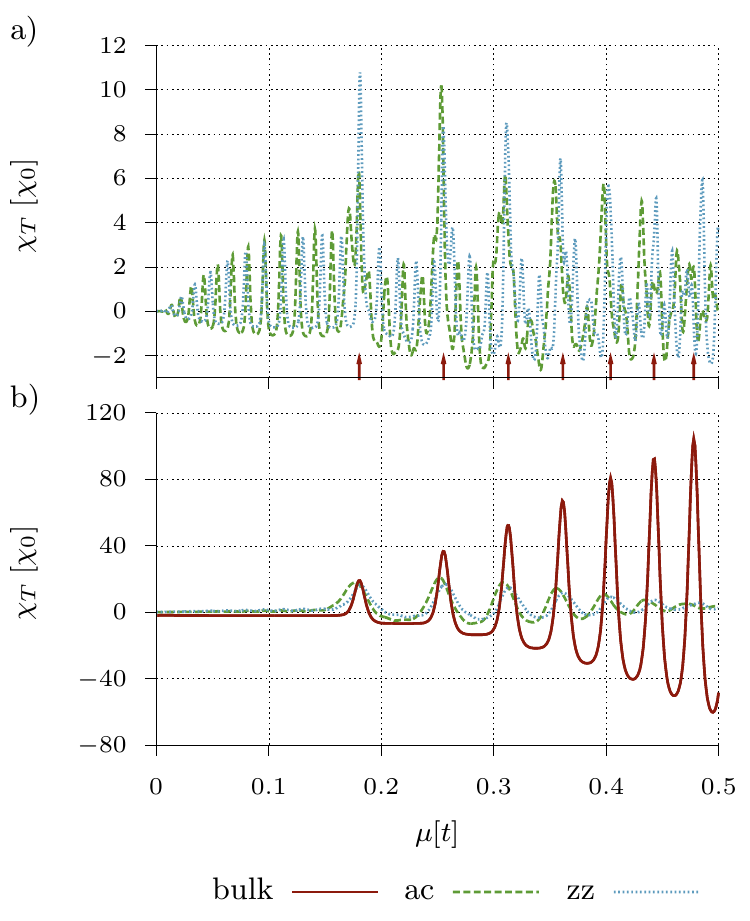}
\caption[]{Oscillatory susceptibility contribution $\chi_{T}$ for triangular graphene cavities as a function of the chemical potential at \mbox{$\phi = 15\,\phi_{0}$}.    
	    a) $\chi_{T}$ for armchair (green dashed) and zigzag (blue dotted) confinement with side length \mbox{$\mathcal{L} \approx 100\,a$} at 
	    \mbox{$1/\beta^{\mathrm{(ac, zz)}} \approx  10^{-3}\,t$}.  The red arrows indicate the peak
positions in the case of bulk graphene where only 
	    Landau levels exist.  
	    b) Comparison of $\chi_{T}$ of the finite systems (dashed and dotted) at a slightly higher thermal energy \mbox{$1/\beta^{\mathrm{(ac, zz)}} = 5\cdot 10^{-3}\,t$} 
	    with the corresponding bulk result (solid) at  \mbox{$1/\beta^{\mathrm{(bulk)}} \approx 3\cdot 10^{-3}\,t$}.  
}
\label{fig:new_norm_regime_b_g_1_g_a_I}
\end{figure}

In \mbox{Fig.~\ref{fig:new_norm_regime_b_g_1_g_a_I}a)} we show the numerically calculated susceptibility contribution $\chi_{T}$ for a triangular armchair and zigzag quantum dot 
for the same value of the magnetic flux as in \mbox{Fig.~\ref{fig:new_norm_regime_b_g_1_g_a_III}a)}, i.e. $\phi = 15\,\phi_{0}$.  The thermal energies are chosen 
as \mbox{$1/\beta^{\mathrm{(ac, zz)}} \approx 1\cdot 10^{-3}\,t$} to satisfy \mbox{relation (\ref{eq:m1_17})}.  
The levels of the finite systems are then well resolved leading to extra peaks with smaller amplitude inbetween 
those caused by level clustering in the vicinity of Landau levels 
(see \mbox{Fig.~\ref{fig:spectrum_qd}}). The latter are indicated by red arrows in \mbox{Fig.~\ref{fig:new_norm_regime_b_g_1_g_a_I}a)} and coincide with the maxima in $\chi_{T}$ of the bulk system.  
These extra peaks are signatures of the confinement of the system and not captured within the bulk theory.  
Similar signatures are numerically observed in \mbox{Ref.~[\onlinecite{PhysRevB.87.115433}]} for
triangular but also hexagonal graphene quantum dots.
In \mbox{Sec. \ref{sec:Main_2}} we will show how one can interprete these finite-size signatures within a semiclassical approach using periodic orbit theory.  
The amplitudes of the susceptibility oscillations of the quantum dots exceed the contribution $\chi_{0}$ from the filled valence band as well, implying that for 
certain ranges of $\phi$ and $\mu$ the total orbital magnetic susceptibility can become paramagnetic.  
By raising the thermal energy to \mbox{$1/\beta^{\mathrm{(ac, zz)}} \approx 5\cdot 10^{-3}\,t$} the
finite-size features are smeared out and only extrema at the positions of the Landau levels survive 
as \mbox{Fig.~\ref{fig:new_norm_regime_b_g_1_g_a_I}b)} demonstrates.  

\begin{figure}[htbp]
 \centering
\includegraphics[width = 0.5\textwidth]{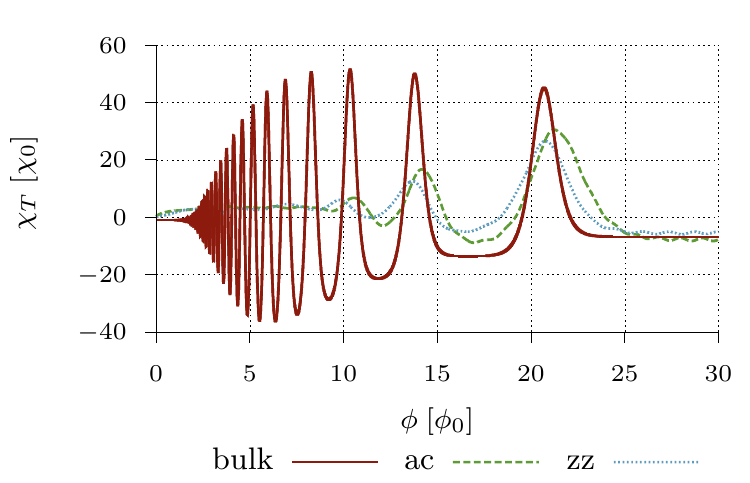}
\caption[]{Susceptibility contribution $\chi_{T}$ as a function of $\phi$ 
(for same triangular quantum dots as in \mbox{Fig.\ \ref{fig:new_norm_regime_b_g_1_g_a_I}})
for \mbox{$\mu = 0.3\,t$, $1/\beta^{\mathrm{(bulk)}} \approx 3\cdot 10^{-3}\,t$} and 
	    \mbox{$1/\beta^{\mathrm{(ac, zz)}} = 5\cdot 10^{-3}\,t$}.  
	    The maxima coincide well with the bulk case for \mbox{$\phi \gtrsim 10\,\phi_{0}$}, i.e.\  the 
  flux range where the bulk theory is applicable to the spectra of the finite quantum dots  
    (see \mbox{Fig.~\ref{fig:spectrum_qd}}).
}
\label{fig:new_norm_regime_b_g_1_g_a_II}
\end{figure}

\mbox{Figure \ref{fig:new_norm_regime_b_g_1_g_a_II}} compares the susceptibility contribution $\chi_{T}$
of the triangular quantum dots with the bulk system as a function of $\phi$ at \mbox{$\mu = 0.3\,t$} and 
\mbox{$1/\beta^{\mathrm{(bulk)}} \approx 3\cdot 10^{-3}\,t$}, respectively \mbox{$1/\beta^{\mathrm{(ac, zz)}} = 5\cdot 10^{-3}\,t$}, such that finize-size effects are smeared out.  
For flux values \mbox{$\phi \gtrsim 10\,\phi_{0}$} the peak positions coincide very well.  
This corresponds to the spectral regime  of the finite systems (see \mbox{Fig \ref{fig:spectrum_qd}})
where the levels cluster in the vicinity of Landau levels and the influence of the boundaries becomes negligible.   

\subsubsection{\label{sec:Main_1_therm_2}Regime: $\alpha, \gamma > 1$}

When the thermal energy and chemical potential are comparable to or smaller than the Landau level spacing 
the temperature dependent part of the susceptibility is 
expected to vanish\cite{PhysRevB.75.115123}.  If the magnetic field is tuned to very high values such that \mbox{$\alpha, \gamma \gg 1$} the degeneracy of each Landau level  
rises accordingly, and all occupied states condense into the first or even to the zeroth level $E_{0}$.  
This yields a contribution to the temperature dependent part of the total grand potential $\Omega_{T}$ linear in $B$ as mentioned in \mbox{Sec. \ref{sec:Main_1_dens}}. 
In order to calculate $\chi_{T}$ from \mbox{Eq.~(\ref{eq:m1_27})} it is useful to apply the representation (\ref{eq:m1_15}) for $\rho^{\mathrm{osc}}$.  
Then it is sufficient to consider only the sum over the Landau indices since the other terms do not contribute to $\chi$.    
To this end we write
\begin{align}
 \label{eq:m1_49}
\tilde{\Omega}_{T}(\mu, B) =\hat{\Omega}_{T}(\mu, B)- g\frac{\varphi}{\beta}\sum\limits_{s = \pm 1}\sum\limits_{n = 1}^{\infty}
	  \ln\left(1 +  \ex^{-\sqrt{2 n}\gamma + s\frac{\gamma}{\alpha}}\right)
\end{align} 
where the $B$-linear term 
\begin{align}
 \label{eq:m1_50}
 \hat{\Omega}_{T}(\mu, B) = 
	   - \frac{g}{2}\frac{\varphi}{\beta}\sum\limits_{s = \pm 1}
	    \ln\left(1 + \ex^{s\frac{\gamma}{\alpha}}\right) - \bar{\Omega}_{T}(\mu)
\end{align}
does not contribute to $\chi_{T}$.  
$\bar{\Omega}_{T}$, defined through \mbox{Eqs.~(\ref{eq:i7},\ref{eq:m1_14})} 
is only based on the average DOS \mbox{$\bar{\rho} = C\left|E\right|$}.  
In order to get an appropriate expression for $\tilde{\Omega}_{T}$ in this parameter range we Taylor expand the
logarithm and the exponential function in \mbox{Eq.~(\ref{eq:m1_49})} using the condition $\gamma \! > \! 1$.  
Resumming the resulting triple infinite sums yields 
\begin{align}
 \label{eq:m1_51}
\begin{split}
 \tilde{\Omega}_{T}(\mu, B) \approx &\,  \hat{\Omega}_{T}(\mu, B) - g \frac{\varphi}{\beta}
		  \sum\limits_{s \pm 1}\mathrm{ln}\left(1+\ex^{-\sqrt{2}\gamma + s\frac{\gamma}{\alpha}}\right),
\end{split}
\end{align}
as shown in \mbox{App.~\ref{sec:App_Int_Omega}}. 
Only the second term contributes to $\chi$.  
It is identical to the contribution from the first electron- and hole-like Landau level to
$\tilde{\Omega}_{T}$ as a comparison with \mbox{Eq.~(\ref{eq:m1_49})} shows.  
The susceptibility contribution from \mbox{Eq.~(\ref{eq:m1_51})} then yields 
\begin{align}
 \label{eq:m1_52}
 \chi_{T}(\mu, B) = & - \frac{1}{8} \sqrt{\frac{\pi}{2}} \frac{\mu_{0} g}{\phi_{0}^{2}}\hbar v_{F} 
\sqrt{\frac{\mathcal{A}}{\varphi}}\times \mathrm{F}\left(\alpha, \gamma\right)\\
 \label{eq:m1_53}
		    = &\, \chi_{0}(B)\times\frac{\pi}{6\sqrt{2}\zeta\left(\frac{3}{2}\right)}\times\mathrm{F}\left(\alpha, \gamma\right).
\end{align}
Here 
\begin{align}
 \label{eq:m1_54}
\begin{split}
 \mathrm{F}\left(\alpha, \gamma\right) = \sum\limits_{s = \pm 1} 
	\left[3\left(1+\ex^{-\sqrt{2}\gamma + s\frac{\gamma}{\alpha}}\right) - \sqrt{2}\gamma\right]\\
	    \times\mathrm{sech}^{2}\left[\frac{1}{2}\left(\sqrt{2}\gamma - s\frac{\gamma}{\alpha}\right)\right]
\end{split}
\end{align}
can assume positive or negative values hence yielding a dia- or paramagnetic susceptibility contribution.  
For \mbox{$\gamma \gtrsim 1$}, i.e. the level spacing is comparable to the thermal energy, $\mathrm{F}\left(\alpha, \gamma\right)$ takes positive values and hence $\chi_{T}$
is diamagnetic.  
In \mbox{Ref.~[\onlinecite{PhysRevB.75.115123}]} the same parameter regime is discussed for the special case $\mu = 0$ but treated in a slightly different way obtaining a 
diamagnetic result for $\chi_{T}$ which decays as a function of $\gamma$.  
In the range of validity of \mbox{Eqs.~(\ref{eq:m1_52}, \ref{eq:m1_53})} $\left|\chi_{T}\right|$ is at most half as large as $|\chi_{0}|$ as the following considerations show:
In its validity range, $\mathrm{F}$ approaches a supremum \mbox{$\lim_{\alpha, \gamma \rightarrow 1}\mathrm{F}(\alpha, \gamma) \approx 4$}.  
Together with the additional prefactors \mbox{$\pi/[6\sqrt{2}\zeta(3/2)] \approx 0.14$} in \mbox{Eq.~(\ref{eq:m1_53})} this yields \mbox{$\chi_{T} \lesssim 0.56 \chi_{0}$}.

\begin{figure}[htbp]
 \centering
\includegraphics[width = 0.5\textwidth]{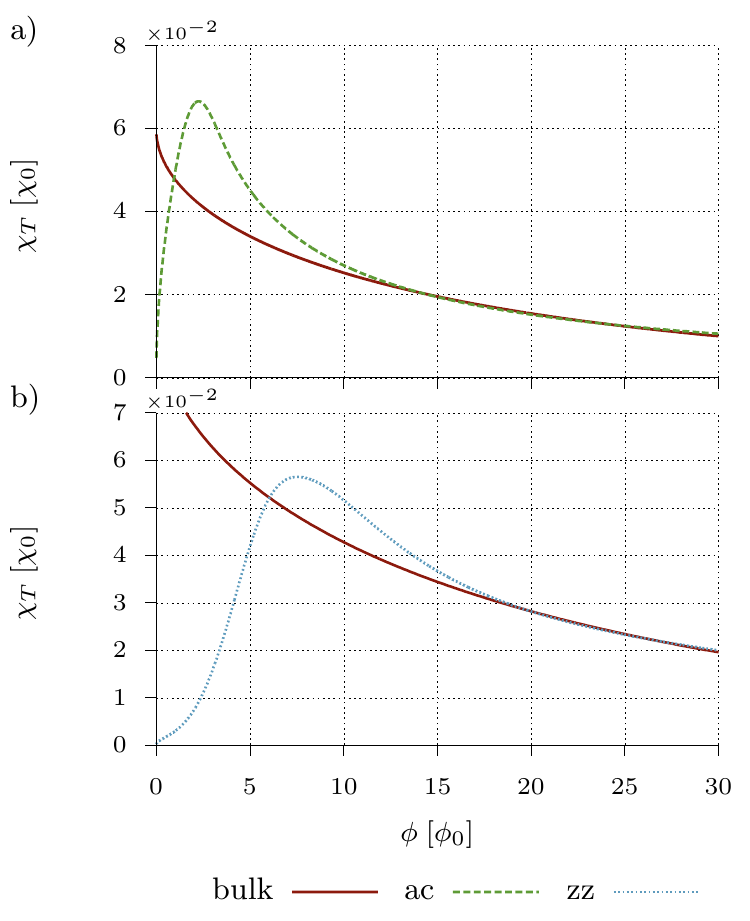}
\caption[]{Flux dependence of the temperature dependent susceptibility contribution $\chi_{T}$ of bulk graphene compared with the numerically calculated contribution of triangular nanostructures with a) 
	    armchair and zigzag b) edges of side length \mbox{$\mathcal{L}\approx 100\,a$}.  The chemical potential is chosen at $\mu = 0$ and the 
	    thermal energies are \mbox{$1/\beta^{\mathrm{(ac, zz)}} \approx 10^{-2}\,t$} and \mbox{$1/\beta^{\mathrm{(bulk)}} \approx 1.5\cdot10^{-1}\,t$} 
	    in panel a) and \mbox{$1/\beta^{\mathrm{(bulk)}} \approx 1.7\cdot10^{-1}\,t$} in panel b), such
that \mbox{Eq.~(\ref{eq:m1_17})} holds true.  
	    The scaling factors \mbox{$\gamma^{\mathrm{(ac)}} \approx 2.5\cdot 10^{-2}$} 
 and \mbox{$\gamma^{\mathrm{(zz)}} \approx 3.8\cdot 10^{-2}$} 
 are obtained by fitting.
}
\label{fig:norm_regime_b_a_g_1_I}
\end{figure}

\begin{figure}[htbp]
 \centering
\includegraphics[width = 0.5\textwidth]{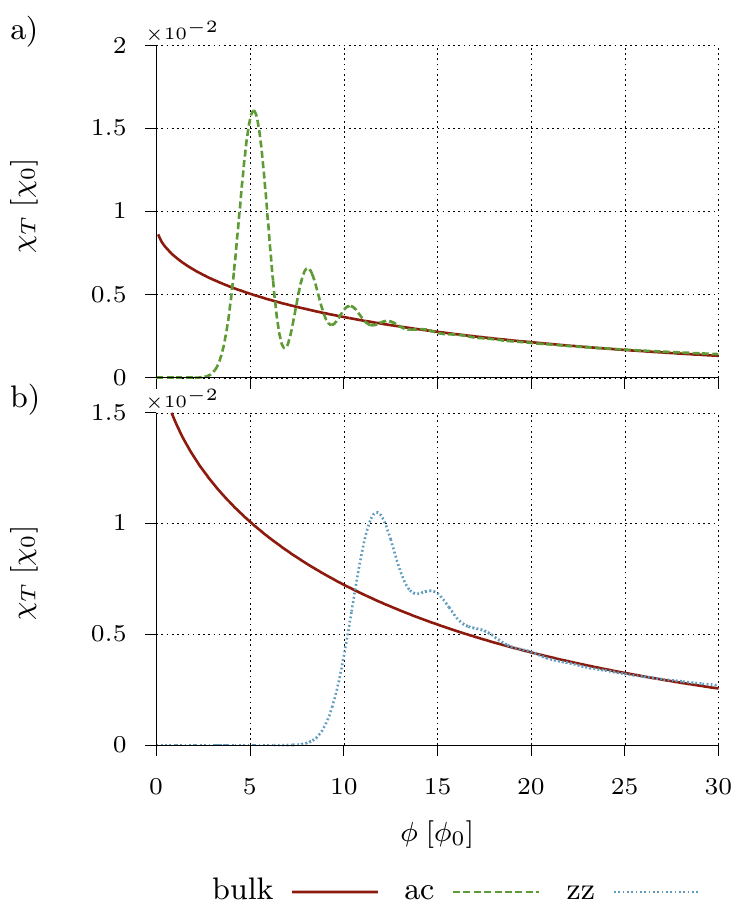}
\caption[]{Same as \mbox{Fig.~\ref{fig:norm_regime_b_a_g_1_I}} for smaller 
	    thermal energies \mbox{$1/\beta^{\mathrm{(ac, zz)}} \approx 10^{-3}\,t$} and \mbox{$1/\beta^{\mathrm{(bulk)}} \approx 1.44\cdot10^{-1}\,t$} 
	    in panel a) and \mbox{$1/\beta^{\mathrm{(bulk)}} \approx 1.42\cdot10^{-1}\,t$} in panel b), such
that \mbox{Eq.~(\ref{eq:m1_17})} holds true.  
	    The scaling factors \mbox{$\gamma^{\mathrm{(ac)}} \approx 3.9\cdot 10^{-3}$} 
 and \mbox{$\gamma^{\mathrm{(zz)}} \approx 7.8\cdot 10^{-3}$} 
 are obtained by fitting.
}
\label{fig:norm_regime_b_a_g_1_II}
\end{figure}

In \mbox{Fig.~\ref{fig:norm_regime_b_a_g_1_I}} the flux dependence of the bulk result, \mbox{Eq.~(\ref{eq:m1_53})}, 
is compared with the numerically calculated contribution from the conduction and valence band to $\chi$ of 
an (a) armchair and (b) zigzag triangular quantum dot at \mbox{$\mu = 0$}.    
The thermal energies of the bulk systems are chosen such that 
\mbox{$\beta^{\mathrm{(bulk)}}\langle\Delta\bar{E}^{\mathrm{(bulk)}}\rangle_{\phi} \approx
\beta^{\mathrm{(ac, zz)}}\langle\Delta\bar{E}^{\mathrm{(ac, zz)}}\rangle_{\phi}$}.
By choosing lower thermal energies finite size effects gain importance and deviations from the bulk theory 
emerge as can be seen from \mbox{Fig.~\ref{fig:norm_regime_b_a_g_1_II}}:  
The susceptibilities $\chi_{T}$ of the quantum dots exhibit oscillatory behavior which becomes all the more pronounced, as the thermal energies tend to lower values.  
In this case all parameters are chosen as in  \mbox{Fig.~\ref{fig:norm_regime_b_a_g_1_I}} but the thermal energy of the quantum dots is one order of magnitude smaller, i.e. \mbox{$1/\beta^{\mathrm{(ac, zz)}} \approx 10^{-3}\,t$}.
For these parameters, the function $\mathrm{F}(\alpha, \gamma)$, \mbox{Eq.~(\ref{eq:m1_54})}, reaches positive values only in the considered flux range.  
Therefore $\chi_{T}$, \mbox {Eq.~(\ref{eq:m1_53})}, is diamagnetic.  
This holds also true for the numerically calculated contribution $\chi_{T}$ of the triangular quantum dots.  
From the definition (\ref{eq:m1_34}) of $\alpha \propto \Delta_{LL}/\mu$ one expects the bulk effects to dominate over finite-size signatures and therefore 
good agreement of the numerical data with the bulk calculations for $\phi \gtrsim 15\ \phi_{0}$.  
This is confirmed by \mbox{Figs.~\ref{fig:norm_regime_b_a_g_1_I}} and \ref{fig:norm_regime_b_a_g_1_II}.

For lower values of $\phi$  \mbox{Eq.~(\ref{eq:m1_52})} is no longer valid yielding deviations from the
tight-binding calculations as the oscillatory modulations of $\chi_{T}$ demonstrate in \mbox{Fig.~\ref{fig:norm_regime_b_a_g_1_II}}.  
These oscillations are smeared out due to the larger thermal energies chosen in \mbox{Fig.~\ref{fig:norm_regime_b_a_g_1_I}}.   
In both figures $\chi_{T}$ of the quantum dots reaches zero for $\varphi \approx 0$ and $\chi_{T}$ is 
morevoer suppressed on a finite flux interval, $\phi \lesssim 3\,\phi_{0}$ in \mbox{Fig.~\ref{fig:norm_regime_b_a_g_1_II}a)} and 
$\phi \lesssim 7\,\phi_{0}$ in b), respectively.  
This behavior can be understood in view of the energy spectra of the quantum dots, \mbox{Fig.~\ref{fig:spectrum_qd}}.  
In each case there is a small gap between \mbox{$E=0$} and the first non-zero energy level as a signature of confinement.
For thermal energies smaller than this gap and \mbox{$\mu = 0$} there are no occupied states above the
Dirac point besides the edge states of the zigzag quantum dot contributing 
$\varphi$-linear to $\Omega_{T}$ and yielding $\chi_{T} = 0$.
In the case of the armchair quantum dot $\Omega_{T}$ and therefore $\chi_{T}$ vanish completely in this specific parameter range.  

\subsubsection{\label{sec:Main_1_therm_3}Regime: $\gamma < 1$ and arbitrary $\alpha$}

If the thermal energy of the system is larger than the level spacing, also states above the Fermi level are occupied implying that tuning the chemical potential or the magnetic field does not lead to a discontinuity 
of the corresponding contribution to the grand potential.  
As a consequence the susceptibility is expected to be a smooth function of these parameters.  
In this parameter range the magnetic flux and the thermal energy can be chosen in such a way that the
Landau level clustering in the quantum dot spectra is pronounced enough to make the bulk theory valid,
 on the one hand, and effectively wash out the finite size signatures on the other hand.    
Hence one can expect good agreement of the bulk theory with the susceptibility of the quantum dots. 

Using again the decomposition of the Fresnel integral into smooth and oscillatory part, one can start
from representation \mbox{Eq.~(\ref{eq:m1_38})} of the field-dependent part of $\Omega$.
Substituting \mbox{$E = 2/\beta\, x + \mu$} gives 
\begin{align}
 \hspace{-0.5cm}\label{eq:m1_55}
\begin{split}
 \tilde{\Omega} =& \frac{K}{2 \sqrt{2}\pi}\varphi^{3/2}\sum\limits_{m=1}^{\infty} \frac{1}{\sqrt{m}^{3}} 
		\mathrm{Im}\!\left[\int\limits_{0}^{\infty}\mathrm{d}u \frac{1}{\sqrt{u}(u - \im)}\right.\\
		&\times\left.\int\limits_{-\infty}^{\infty}\mathrm{d}x\, \mathrm{sech}^{2}\left(x\right)
		\ex^{-u \pi m \left(2 \frac{x}{\gamma} + \frac{1}{\alpha}\right)^{2}}\ex^{ \im \pi m \left(2 \frac{x}{\gamma} + \frac{1}{\alpha}\right)^{2}}\right].\hspace{-0.5cm}
\end{split}
\end{align}
For $\gamma < 1$ the complex phase rapidly oscillates as a function of $x$ for all values of $\alpha$.  
Therefore the second integral 
can be solved within stationary phase approximation:
\begin{equation}
\label{eq:m1_56}
 \int\limits_{-\infty}^{\infty}\!\!\mathrm{d}x\, \mathrm{sech}^{2}\left(x\right)
		\ex^{\im \pi m \left(2 \frac{x}{\gamma} + \frac{1}{\alpha}\right)^{2}} \!
		\approx \frac{\left|\gamma\right|}{2\sqrt{m}}\,\mathrm{sech}^{2}\left(\frac{\gamma}{2 \alpha}\right) \ex^{\im\frac{\pi}{4}}.
\end{equation}
Using \mbox{$\int_{0}^{\infty}\mathrm{d}u [\sqrt{u}(u-\im)]^{-1} \!=\! \pi\exp\left(-\im \pi/4\right)$} 
in \mbox{(\ref{eq:m1_55})} yields 
\begin{align}
\label{eq:m1_57}
 \tilde{\Omega}\approx & \frac{K}{4 \sqrt{2}}{\sqrt{\varphi}^{3}}\left|\gamma\right|\,\mathrm{sech}^{2}\left(\frac{\gamma}{2 \alpha}\right)
		\sum\limits_{m=1}^{\infty} \frac{1}{m^{2}} \\
\label{eq:m1_58}
	      =& \frac{ g}{\mathcal{A}} (\hbar v_{F})^{2}\frac{\pi^{2}}{12}\,\beta \,\mathrm{sech}^{2}\left(\frac{\gamma}{2 \alpha}\right)\,\varphi^{2},		
\end{align}
where \mbox{$\sum_{m=1}^{\infty}m^{-2} = \pi^{2}/6$} is used.  
The corresponding expression for the total orbital susceptibility reads
\begin{align}
\label{eq:m1_59}
 \chi(\mu) =&-\frac{\mu_{0} g}{\phi_{0}^{2}} (\hbar v_{F})^{2} \frac{\pi^{2}}{6} \,\beta \,\mathrm{sech}^{2}\left(\frac{\mu\,\beta}{2}\right)\\
 \label{eq:m1_60}
	=& \chi_{0}(B)\times \frac{\sqrt{2}\pi^{2}}{9\zeta\left(\frac{3}{2}\right)}\,\gamma\,\mathrm{sech}^{2}\left(\frac{\gamma}{2\alpha}\right).
\end{align}

\begin{figure}[htbp]
 \centering
\includegraphics[width = 0.5\textwidth]{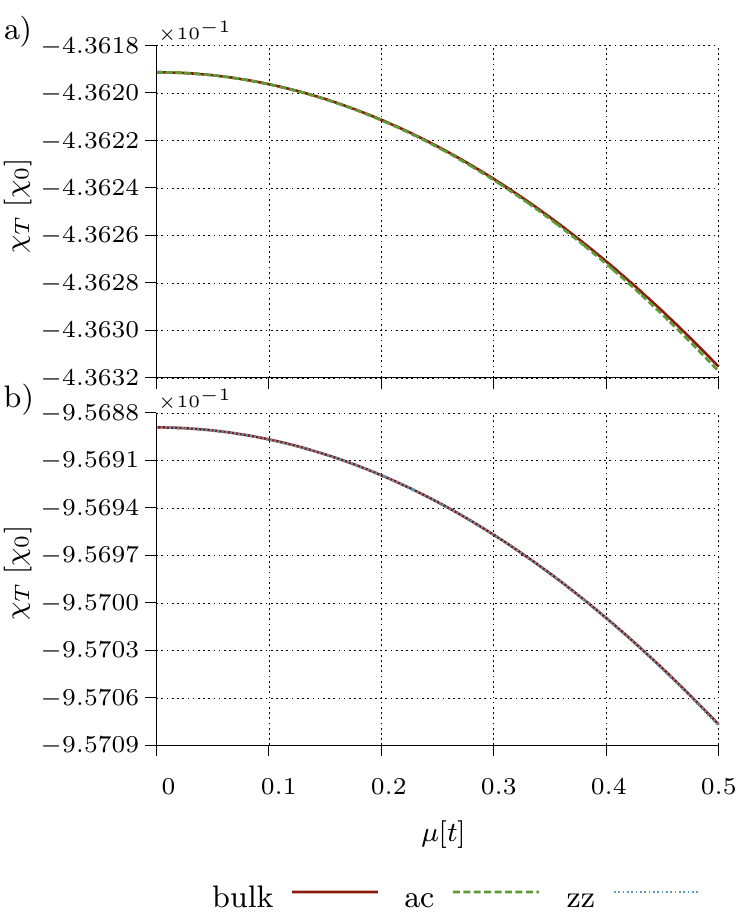}
\caption{Comparison of the temperature-dependent susceptibility contribution $\chi_{T}$ 
for bulk graphene with that of a triangular armchair a) and zigzag b) graphene flake as a function 
	      of the chemical potential at a magnetic flux of $\phi = 5\,\phi_{0}$ and
$1/\beta^{(\mathrm{ac, zz})} = 5\,t$.  
 The corresponding values used in the analytic expression for the bulk are 
	      $1/\beta^{(\mathrm{bulk})} \approx 2.1\,t$ 
      in a) and $1/\beta^{(\mathrm{bulk})} \approx 2.9\,t$
      in b). The fitted scaling factors are $\gamma^{(\mathrm{ac})} = 0.44, 
        \gamma^{(\mathrm{zz})} = 1.7$.}
\label{fig:norm_regime_b_k_1_I}
\end{figure}

In this regime the divergent contribution $\chi_{0}$ of the filled valence band is compensated by the
contribution $\chi_{T}$ of the thermally excited charge carriers leading to a distinctly diamagnetic and
moreover flux independent magnetic response.  
This result can also be found in the literature\cite{PhysRev.104.666, PhysRevB.20.4889, PhysRevB.75.235333, PhysRevB.75.115123}.   
The contribution \mbox{$\chi_{T} = \chi - \chi_{0}$} can be extracted from Eq.~(\ref{eq:m1_60}) reading 
\begin{align}
 \label{eq:m1_61}
 \hspace{-0.5cm}\chi_{T}(\mu)=  -\chi_{0}(B)\, \left[1 -
\frac{\sqrt{2}\pi^{2}}{9\zeta\left(\frac{3}{2}\right)}\,\gamma\,\mathrm{sech}^{2}\left(\frac{\gamma}{2\alpha}\right)\right]
\, .
\end{align}
Since \mbox{$\sqrt{2}\pi^{2}/[9\zeta(3/2)] \approx   0.6$} and \mbox{$\mathrm{sech}^{2}(x) \leq 1$}, \mbox{$\forall x\in\mathbb{R}$} 
the contribution $\chi_{T}$ exhibits paramagnetic behavior in this parameter range.  
The comparison of this bulk contribution with numerical data for the triangular armchair and zigzag quantum dot 
in \mbox{Fig.~\ref{fig:norm_regime_b_k_1_I} a)} and b) shows perfect agreement as expected at larger fluxes.  

To fulifil $\gamma < 1$, i.e. $\sqrt{\mathcal{A}/(2 \pi)\,\varphi} < k_{B} T/(\hbar v_{F})$,
in the limit of very low temperatures requires that 
\mbox{$|E_{n}|$}, \mbox{Eq.~(\ref{eq:landau})}, tend to zero even for large Landau indices $n$.  
Hence a change in the magnetization of bulk graphene due to weakly thermally excited charge carriers can only occur for Fermi energies close to the Dirac point.  
For \mbox{$T\rightarrow 0$} 
this leads to a sharply peaked susceptibility at $\mu = 0$.  
In view of \mbox{Eq.~(\ref{eq:m1_31})}, this can be deduced from \mbox{Eq.~(\ref{eq:m1_59})} yielding the well known expression
\cite{PhysRev.104.666, PhysRevB.20.4889, PhysRevB.75.235333, PhysRevB.76.113301, PhysRevLett.102.177203, JPSJ.80.114705, 1751-8121-44-27-275001, PhysRevB.83.235409,  PhysRevB.80.075418}
\begin{align}
 \label{eq:m1_62}
 \chi(\mu) \xrightarrow{\beta \rightarrow \infty}-\frac{\mu_{0} g}{\phi_{0}^{2}} (\hbar v_{F})^{2} \frac{2 \pi^{2}}{3} \,\delta\left(\mu\right).
\end{align}
This limit is not truly reachable numerically for the finite systems considered 
since the Landau level structure 
is not pronounced enough as it can be seen from \mbox{Fig.~\ref{fig:spectrum_qd}}.    

Another limit of physical relevance concerns $\mu\rightarrow 0$ or $\alpha \rightarrow \infty$.  In this limit the total orbital susceptibility reads  
\begin{equation}
 \label{eq:m1_63}
 \chi(\mu) \xrightarrow{\mu \rightarrow 0}-\frac{\mu_{0} g}{\phi_{0}^{2}} (\hbar v_{F})^{2}
\frac{\pi^{2}}{6} \,\beta \propto - \frac{1}{k_{B} T} .
\end{equation}

\begin{figure}[htbp]
 \centering
\includegraphics[width = 0.5\textwidth]{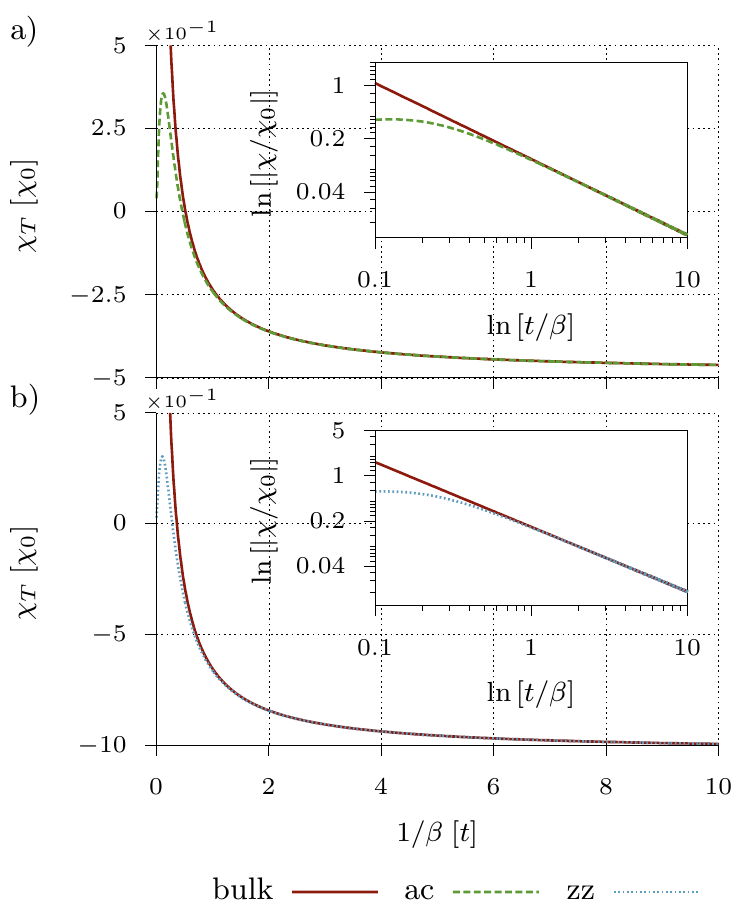}
\caption{Comparison of the numerical data of the orbital magnetic susceptibility contribution $\chi_{T}$ of a triangular armchair a) and zigzag  b) graphene flake  at $\phi = 5\,\phi_{0}$ 
with the analytic result for bulk graphene in the limit \mbox{$\mu \rightarrow 0$}.  In both cases the correspondence is convincing for \mbox{$t/\beta > 1$} as for lower thermal energies this 
approximation loses validity.  The scaling factors attain $\gamma^{(\mathrm{ac})} \approx 5.8$ and $\gamma^{(\mathrm{zz})} \approx 8.6$ which is in agreement with the condition 
\mbox{$|\bar{E}^{(\mathrm{bulk})} - \bar{E}^{(\mathrm{ac})}| < |\bar{E}^{(\mathrm{bulk})} - \bar{E}^{(\mathrm{zz})}|$}.  The insets show both the full orbital magnetic susceptibility $\chi$ 
in a double logarithmic plot and confirm the scaling behavior $\chi \propto -\beta$ at the Dirac point.
}
\label{fig:norm_regime_b_k_1_II}
\end{figure}

This typical temperature dependence, already known in the literature\cite{PhysRevB.75.115123}, is also affirmed by the numerical data, see \mbox{Fig.~\ref{fig:norm_regime_b_k_1_II}}.  
The double logarithmic graphs in the insets show clearly the $1/k_B T$ dependence 
in the limit $\mu\rightarrow 0$ (for $\phi = 5\,\phi_{0}$).  
The difference between the bulk theory and the numerical data for small thermal energies reflects, 
on the one hand, the limit of validity of the analytical approximation for $\gamma < 1$;   
on the other hand, it is a signature of finite-size effects which gain importance in the low temperature limit.  

\section{\label{sec:Main_2}Oscillatory finite-size effects for graphene nanostructures}
\subsection{General semiclassical framework}

\begin{figure}[htbp]
 \centering
\includegraphics[width = 0.42\textwidth]{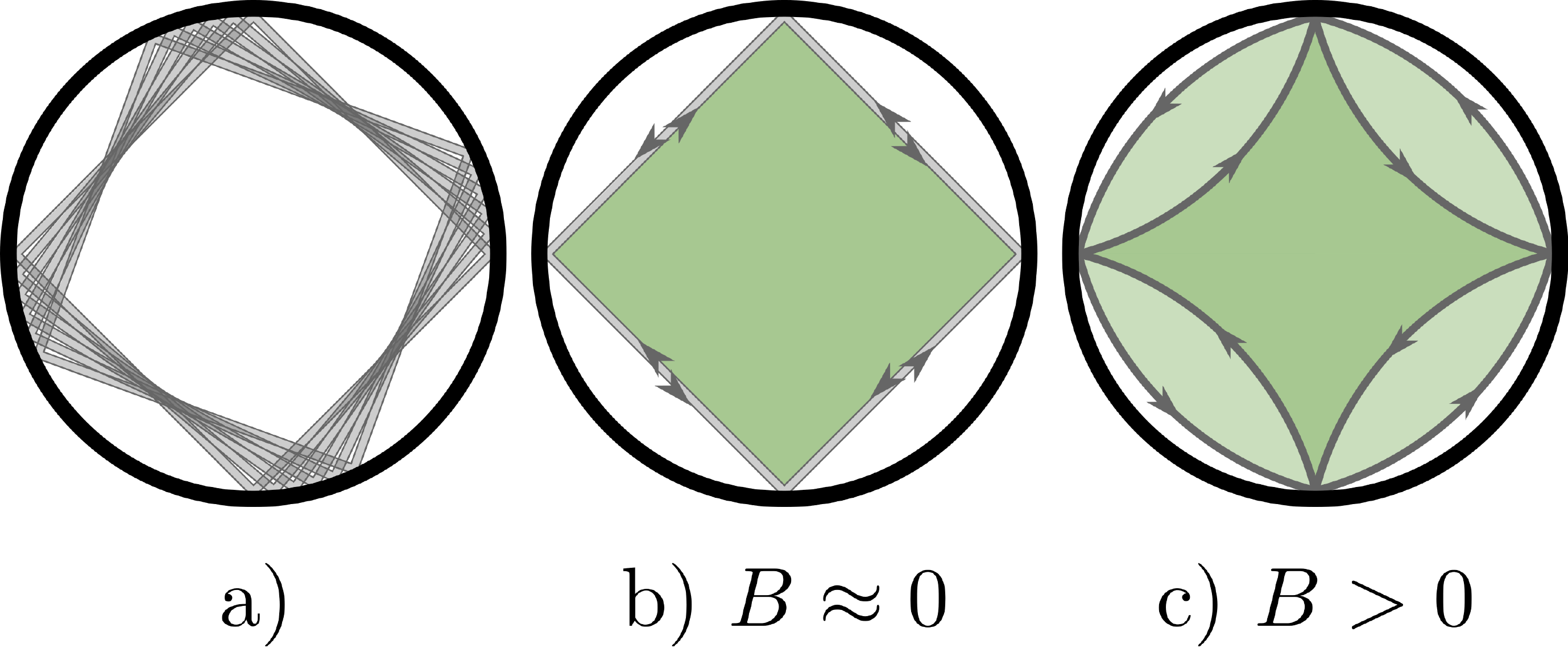}
         \caption{Classical periodic orbits in the circular billiard.
Panel a) shows representatives of one fundamental orbit family.  
Panel b) and c): Pairs of counter-propagating orbits in the presence of a perpendicular magnetic field.  
For weak magnetic fields, panel b), the bending of the classical orbits can be neglected and
the enclosed areas (green shaded) are approximately equal.  
Panel c) shows the same orbit pair for stronger magnetic field.}
\label{fig:TRS}
\end{figure}

Semiclassical periodic orbit theories 
offer a distinguished way to analytically describe finite-size effects encoded in the energy spectra of spatially confined systems of arbitrary shape.  
Boundary effects are incorporated in the semiclassical approximation of the oscillatory part of the DOS, $\rho^{\mathrm{osc}}_{\mathrm{sc}}(E)$.  
One important criteria for applying such semiclassical approximations,
\textit{ the Gutzwiller trace formula}\cite{gutzwiller1990chaos} for chaotic classical dynamics or the 
\textit{ Berry-Tabor trace formula}\cite{0305-4470-10-3-009} for regular classical dynamics,
requires that the linear system size lies in a mesoscopic regime, \mbox{$k\mathcal{L} \gg 1$}, where $k = E/(\hbar v_{F})$ is the Fermi wave number.  
In general, $d^{\mathrm{osc}}_{\mathrm{sc}}(E)$ is of the form
\begin{align}
 \label{eq:sc_64}
 d^{\mathrm{osc}}_{\mathrm{sc}}(E) &= \sum\limits_{\gamma}d^{\mathrm{osc}}_{\mathrm{sc}, \gamma}(E),\\
 \label{eq:sc_64a}
  d^{\mathrm{osc}}_{\mathrm{sc}, \gamma}(E)&\propto \mathrm{Re} D_{\gamma} \ex^{\frac{\im}{\hbar}S_{\gamma}},
\end{align}
where the sum runs over infinitely many classical periodic orbits $\gamma$ with classical action \mbox{$S_{\gamma} = \oint_{\gamma}\mathrm{d}\bold{q}\cdot\bold{p} = p \mathcal{L}_{\gamma}$} 
and length $\mathcal{L}_{\gamma}$.  
The exact form of the classical amplitude $D_{\gamma}$ sensitively depends on the specific geometry of the system and can be calculated either within the recipe given by Gutzwiller\cite{gutzwiller1990chaos} 
in the case of non-integrable classical dynamics 
or within the recipe of Berry and Tabor\cite{0305-4470-10-3-009} when the classical dynamics is integrable.  
In the latter  case, relevant in the following, the summation over $\gamma$ in \mbox{Eq.~(\ref{eq:sc_64})}
runs over families of degenerate orbits, as depicted in \mbox{Fig.~\ref{fig:TRS}a)} for a disk geometry.  
This degeneracy of orbits in a regular billiard can be described in terms of continuous symmetry groups
$G$ such that the members of a specific orbit family are related to each other through the action of a group element $g$ of $\mathbb{G}$.  
This is already included in the Berry-Tabor trace formula\cite{0305-4470-10-3-009} for field-free regular systems.
In the case of small symmetry breaking, as it is caused by an weak external magnetic field, one has to take these degeneracies separately into account as discussed in 
\mbox{Subsec.~\ref{sec:Main_2_trace}}. 
Therefore, we will associate an orbit family $\gamma$ with the corresponding element $g$ of
the underlying symmetry group $\mathbb{G}$ if necessary, i.e. \mbox{$\gamma \mapsto \gamma(g)$}.  

In \mbox{Refs.~[\onlinecite{epub12143, PhysRevB.84.075468, PhysRevB.84.205421}]} the authors show in a general way, how the trace formulas for \"{}Schr\"odinger billiards\"{} with 
classically regular or chaotic dynamics can be extended to an arbitrary shaped, field-free graphene flake including the 
most common types of boundaries, i.e. zigzag, armchair and infinite-mass-type edges.  
Resembling \mbox{Eq.~(\ref{eq:sc_64})} the semiclassical trance fromulas for graphene read
\begin{align}
 \label{eq:sc_65}
  \rho^{\mathrm{osc}}_{\mathrm{sc}}(E) = \sum\limits_{\gamma}\rho^{\mathrm{osc}}_{\mathrm{sc}, \gamma}(E),\quad \rho^{\mathrm{osc}}_{\mathrm{sc}, \gamma}(E)\propto d^{\mathrm{osc}}_{\mathrm{sc, \gamma}}(E)\mathrm{Tr}K_{\gamma},
\end{align}
where $d^{\mathrm{osc}}_{\mathrm{sc, \gamma}}$ is given by \mbox{Eq.~(\ref{eq:sc_64})}, of the corresponding Schr\"odinger system.  
Hence, the $d^{\mathrm{osc}}_{\mathrm{sc}, \gamma}$ contain all information about the orbital dynamics in the graphene system.  
The additional factor $\mathrm{Tr}K_{\gamma}$ denotes a trace over the pseudospin propagator $K_{\gamma}$ of the orbit $\gamma$ and contains only graphene specific information about the boundary.  
In \mbox{Refs.~[\onlinecite{epub12143}, \onlinecite{PhysRevB.84.205421}]} a general expression for $\mathrm{Tr}K_{\gamma}$ of an orbit, with $N_{\gamma}$ reflections at the boundaries is derived, yielding 
\begin{align}
 \label{eq:sc_66} 
  \mathrm{Tr}K_{\gamma} = 4 f_{\gamma}\cos\left(\theta_{\gamma} + \frac{\pi}{2}N_{\gamma}\right)\cos\left(2 K \Lambda_{\gamma} + \vartheta_{\gamma} + \frac{\pi}{2}N_{\gamma}\right),
\end{align}
if the total number of reflections on armchair edges, $N_{\mathrm{ac}}$, is even and $\mathrm{Tr}K_{\gamma} = 0$ otherwise.  
The prefactor is defined as \mbox{$f_{\gamma} = \im^{3 N_{\gamma} - N_{\mathrm{zz}}}$}, where  $N_{\mathrm{zz}}$ denotes the number of reflections on zigzag edges.  
\mbox{$\theta_{\gamma} = \sum_{i = 1}^{N_{\gamma}}\theta_{i}$} is the sum over all reflection angles along the orbit $\gamma$.  
\mbox{$K = 4 \pi/(3 a)$} denotes the distances between the Dirac points and the $\Gamma$ point of the Brillouin zone.  
\mbox{$\Lambda_{\gamma} = \sum_{i = 1}^{N_{\mathrm{ac}}/2}(x_{2 i - 1} - x_{2 i})$} is the sum over the distance between two subsequent reflections on armchair edges.  
Further \mbox{$\vartheta_{\gamma} = \sum_{i = 1}^{N_{\mathrm{zz}}}(-1)^{s_{i}}\vartheta_{i}$} denotes the sum over zz reflection angles $\vartheta_{i}$, 
where \mbox{$\vartheta_{i} = \pm \theta_{i}$} for reflection on A- and B-edges, respectively,  
and $s_{i}$ is the number of 
ac reflections occuring after the zz reflection $i$.  
One finds\cite{epub12143, PhysRevB.84.205421} \mbox{$\mathrm{Tr}K_{\gamma} = \mathrm{Tr}K_{\gamma^{-1}}$}
where $\gamma^{-1}$ denotes the time reversed partner of orbit $\gamma$, 

\subsection{\label{sec:Main_2_trace}Semiclassical approximation of the orbital magnetic susceptibility}

In \mbox{Ref.~[\onlinecite{Richter}]} the authors showed how the semiclassic theory of integrable and non-integrable billiard systems with parabolic dispersion can 
be extended to include the effect of an homogeneous, constant magnetic field.  
Due to the formal similarity of the trace formulas of systems with parabolic dispersion, \mbox{Eq.~(\ref{eq:sc_64})}, and graphene, \mbox{Eq.~(\ref{eq:sc_65})}, 
the techniques used in \mbox{Ref.~[\onlinecite{Richter}]} can be readily transferred.
We will focus on the low-field regime where the classical 
cyclotron radius \mbox{$R_{c} = k\,l_{B}^{2}$} is much larger than the linear system size, i.e.  \mbox{$R_{c} \gg \mathcal{L}$}.  
In the following we will consider quantum dots with corresponding regular classical dynamics in the field-free 
case. A derivation of orbital magentic properties of cavities with chaotic underlying dyanmics can be
derived correspondingly.
Following \mbox{Refs.~[\onlinecite{ StephenC199660}, \onlinecite{bohigas}]}, we treat the weak magnetic field perturbatively, such that the classical Hamiltonian of the system, 
\begin{align}
 \label{eq:sc_67}
 \mathcal{H} = \frac{\left[\bold{p} - e \bold{A}\left(\bold{q}\right)\right]^{2}}{2 m} + V\left(\bold{q}\right),
\end{align}
can be decomposed into the unperturbed part,
\mbox{$\mathcal{H}_{0} = \bold{p}^{2}/(2m) + V\left(\bold{q}\right)$},
and the small perturbation
$  -\frac{1}{m}\bold{p}\cdot\bold{A}\left(\bold{q}\right)$.
To leading perturbative order 
the action difference between an orbit in the perturbed and the unperturbed system reads
n\cite{bohigas, StephenC199660}
\begin{align}
 \label{eq:sc_70}
 \delta S_{\gamma} \approx e \int_{\gamma}\mathrm{d}\bold{q}\cdot\bold{A}\left(\bold{q}\right) =  e \bold{B}\cdot\bold{\mathcal{A}}_{\gamma},
\end{align}
with $\bold{\mathcal{A}}_{\gamma}$ the directed, enclosed area of the unperturbed orbit $\gamma$.  
In \mbox{Refs.~[ \onlinecite{Richter}, \onlinecite{brack_sc}, \onlinecite{StephenC199660}]} it is 
moreover shown that in the presence of a weak magnetic field 
the trace formula (\ref{eq:sc_64}) for the field-free Schr\"odinger system is modified to 
\begin{align}
 \label{eq:sc_71}
\begin{split}
 d^{\mathrm{osc}}_{\mathrm{sc}}(E, B) =& \sum\limits_{\gamma}d^{\mathrm{osc}}_{\mathrm{sc}, \gamma}(E, B),\\
  d^{\mathrm{osc}}_{\mathrm{sc}, \gamma}(E, B)\propto&\,\mathrm{Re}\left[D_{\gamma} \ex^{\frac{\im}{\hbar}S_{0, \gamma}}\times\mathcal{M}_{\gamma}(B)\right],
\end{split}
\end{align}
with the field-dependent modulation factor
\begin{align}
 \label{eq:sc_72}
 \mathcal{M}_{\gamma}(B) = \frac{1}{V_{g}}\!\int_{\mathbb{G}}\!\mathrm{d}\mu\!\left(g\right)\ex^{\frac{\im}{\hbar}\delta S_{\gamma(g)}}
	  = \frac{1}{V_{g}}\!\int_{\mathbb{G}}\!\mathrm{d}\mu\!\left(g\right)\ex^{\im\frac{2\pi}{\phi_{0}}\bold{B}\cdot\bold{\mathcal{A}}_{\gamma(g)}}.
\end{align}
The index $g$ represents an element of the symmetry group $\mathbb{G}$ characterizing the degeneracy of orbits $\gamma(g)$ in one specific orbit family.  
Since $\mu(g)$ is the Haar measure\cite{Haar} of $\mathbb{G}$, the normalization factor \mbox{$V_{g} = \int_{\mathbb{G}}\mathrm{d}\mu(g)$} can be understood as the volume of $\mathbb{G}$.
Since $d^{\mathrm{osc}}_{\mathrm{sc}}$ contains all information of the orbital dynamics, including the 
influence of the $B$-field, we can adapt \mbox{Eq.~(\ref{eq:sc_71})} and derive the oscillatory part 
of the DOS for a regular graphene cavity in a weak magnetic field in semiclassical approximation:
\begin{align}
 \label{eq:sc_72b}
\begin{split}
 \rho^{\mathrm{osc}}_{\mathrm{sc}}(E, B) =& \sum\limits_{\gamma}\rho^{\mathrm{osc}}_{\mathrm{sc},\gamma}(E, B),\\
 \rho^{\mathrm{osc}}_{\mathrm{sc},\gamma}(E, B) \propto&\, d^{\mathrm{osc}}_{\mathrm{sc},\gamma}(E, B)\mathrm{Tr}K_{\gamma}.
\end{split}
\end{align}
\mbox{Equation (\ref{eq:sc_72b})} is applicable to both, systems that remain integrable in a weak magnetic
field and systems which are no longer integrable due to the symmetry breaking caused by a weak magnetic field, 
e.g. a rectangular quantum dot considered in \mbox{Subsec.~\ref{sec:Main_2_rect}}.  
The lengths of time-reversed partner orbits or families, $\gamma$ and $\gamma^{-1}$, (for $B\!=\!0$) 
are equal, but the directed, enclosed areas have opposite signs due to the propagation direction, i.e. 
\mbox{$\mathcal{L}_{\gamma} = \mathcal{L}_{\gamma'}$} and 
\mbox{$\mathcal{A}_{\gamma} = -\mathcal{A}_{\gamma'}$}. 
The contribution of these orbit pairs to the DOS can be combined to
\begin{align}
 \label{eq:sc_72ba}
 \rho^{\mathrm{osc}}_{\mathrm{sc}, \gamma}(E, B) +& \rho^{\mathrm{osc}}_{\mathrm{sc}, \gamma'}(E, B) = 2 \,\rho^{\mathrm{osc}}_{\mathrm{sc}, \gamma}(E)\times\mathcal{C}_{\gamma}(B),
\end{align}
where $\rho^{\mathrm{osc}}_{\mathrm{sc}, \gamma}(E)$ is the contribution (\ref{eq:sc_65}) of the orbit family $\gamma$ to $\rho^{\mathrm{osc}}_{\mathrm{sc}}$ in the field-free 
system and 
\begin{align}
 \label{eq:sc_72c}
 \mathcal{C}_{\gamma}(B) = \frac{1}{V_{g}}\!\int_{\mathbb{G}}\!\mathrm{d}\mu\!\left(g\right)\cos\left(\frac{2\pi}{\phi_{0}}\bold{B}\cdot\bold{\mathcal{A}}_{\gamma(g)}\right).
\end{align}
The field dependence of the DOS and therefore of related observables such as the magnetic susceptibility is governed by dephasing between time-reversed orbit families 
and affected by dephasing between different members of a given orbit family induced by the magnetic field.  
From definition (\ref{eq:m1_28}) of the grand  potential one can 
deduce the semiclassical approximation of the oscillatory part\cite{Richter}
\begin{align}
 \label{eq:sc_73}
 \Omega^{\mathrm{osc}}_{\mathrm{sc}}(\mu, B) =& \int\limits_{-\infty}^{\infty}\mathrm{d}E\,\mathcal{N}^{\mathrm{osc}}_{\mathrm{sc}}(E, B) f'(E - \mu),
\end{align}
where $\mathcal{N}^{\mathrm{osc}}_{\mathrm{sc}}$ is obtained from $\rho^{\mathrm{osc}}_{\mathrm{sc}}$ after 
integrating twice by parts.
For the contribution of the orbit family $\gamma$ to the oscillatory DOS, $\rho^{\mathrm{osc}}_{\mathrm{sc}, \gamma}$, one finds\cite{Richter}
\begin{align}
 \label{eq:sc_74}
 \mathcal{N}^{\mathrm{osc}}_{\mathrm{sc}, \gamma}(E, B) = -\left(\frac{\hbar}{\mathrm{d}S_{\gamma}/\mathrm{d}E}\right)^{2} \rho^{\mathrm{osc}}_{\mathrm{sc}, \gamma}(E, B).
\end{align}
The energy integral (\ref{eq:sc_73}) is of the form of \mbox{Eq.~(\ref{eq:m1_39})} and solved as described
in App.\ A of \mbox{Ref.~[\onlinecite{Richter}]}.
Using \mbox{Eq.~(\ref{eq:m1_41})} and  
\mbox{$\mathrm{d}S_{\gamma}/\mathrm{d}E =\tau_{\gamma} = \mathcal{L}_{\gamma}/v_{F}$} one eventually finds 
\begin{align}
 \label{eq:sc_75}
\hspace{-0.4cm} \Omega^{\mathrm{osc}}_{\mathrm{sc}}(\mu, B) \approx \sum\limits_{\gamma} \left(\frac{\hbar v_{F}}{\mathcal{L}_{\gamma}}\right)^{2}\!\!\!\rho^{\mathrm{osc}}_{\mathrm{sc}, \gamma}(\mu, B)
 \mathrm{R}_{T}\left(\frac{\mathcal{L}\gamma}{\mathcal{L}_c}\right).\hspace{-0.3cm}
\end{align}
At finite $T$ the sum converges due to the exponential suppression of orbit families  with 
\mbox{$\mathcal{L}_{\gamma} > \mathcal{L}_{c} = \hbar v_{F}\,\beta/\pi$} encoded in $\mathrm{R}_{T}$, \mbox{Eq.~(\ref{eq:m1_42})}.   
Taking twice the $B$-field derivative one finds the semiclassical, oscillatory contribution to the
orbital susceptibility 
of a graphene nanostructure with underlying regular classical dynamics:
\begin{align}
 \label{eq:sc_76}
\begin{split}
 \chi^{\mathrm{osc}}_{\mathrm{sc}}(\mu, B) =& -\frac{\mu_{0}}{\mathcal{A}}
	\sum\limits_{\gamma}  \left(\frac{\hbar
v_{F}}{\mathcal{L}_{\gamma}}\right)^{2}\mathrm{R}_{T}\left(\frac{\mathcal{L}_{\gamma}}{\mathcal{L_c}}\right)\\
&\times f_{\gamma}\,\rho^{\mathrm{osc}}_{\mathrm{sc}, \gamma}(\mu)\frac{\partial^{2}}{\partial
B^{2}}\mathcal{C}_{\gamma}(B) \, .
\end{split}
\end{align}
Here, the sum involves one propagation direction of orbit families $\gamma$.  
Time-reversed partners are considered by the factor $f_{\gamma}\!=\!2$.
The magnetic phase factor $\mathcal{C}_{\gamma}$, \mbox{Eq.~(\ref{eq:sc_72c})}, implies that only 
orbits contribute to $\chi^{\mathrm{osc}}_{\mathrm{sc}}$ that enclose a finite area in the field-free
case, and hence  self-retracing orbits ($f_{\gamma} \!=\! 1$) do not contribute.
We note that the same formal expression (\ref{eq:sc_76}) holds true for Schr\"odinger-type systems and graphene, 
since the graphene-specific relevant information is implicitly contained in 
$\rho^{\mathrm{osc}}_{\mathrm{sc, \gamma}}$.  

In the following we compare these predictions for the orbital magnetic response
with quantum mechanical results within the effective Dirac model 
(\mbox{Subsec.~\ref{sec:Main_2_disk}}) 
and full tight-binding calculations (\mbox{Subsec.~\ref{sec:Main_2_rect}}).

\subsection{\label{sec:Main_2_disk}Circular billiard with infinite-mass-type edges}

The first representative system we analyze is a disk-shaped graphene quantum dot with infinite-mass-type edges.  
Due to its rotational symmetry there is a separable quantum mechanical solution within the Dirac
approximation even in the presence of a magnetic field. The resulting quantization condition 
reads\cite{BM, PhysRevB.76.235404}
\begin{align}
 \label{eq:sc_d1}
 J_{\bar{m}}\left(k_{\bar{m}n}R\right) = \tau J_{\bar{m} + 1}\left(k_{\bar{m}n}R\right) \, .
\end{align}
Here, $\tau = \pm 1$ labels the two valleys of the graphene Brillouin zone, $R$ is the disk radius
and $J_{v}(x)$ denotes the $v$-th order Bessel function of the first kind\cite{gradshtein}.
The index $\bar{m} = m + \phi/\phi_{0}$ includes the magnetic flux $\phi$ and the azimuthal 
orbital angular momentum quantum number \mbox{$m = 0, \pm 1, ...$} .
The second quantum number $n \in \mathbb{Z}$ counts (for a given $\bar{m}$) the solutions $k_{\bar{m}n}$
to \mbox{Eq.~(\ref{eq:sc_d1})} which are obtained numerically.  
Each energy level has a two fold spin degeneracy. 
Based on \mbox{Eq.~(\ref{eq:sc_d1})}, one can calculate the orbital magnetic susceptibility quantum mechanically according to \mbox{Eq.~(\ref{eq:i6})}.  

\begin{figure}[htbp]
 \centering
\includegraphics[width = 0.45\textwidth]{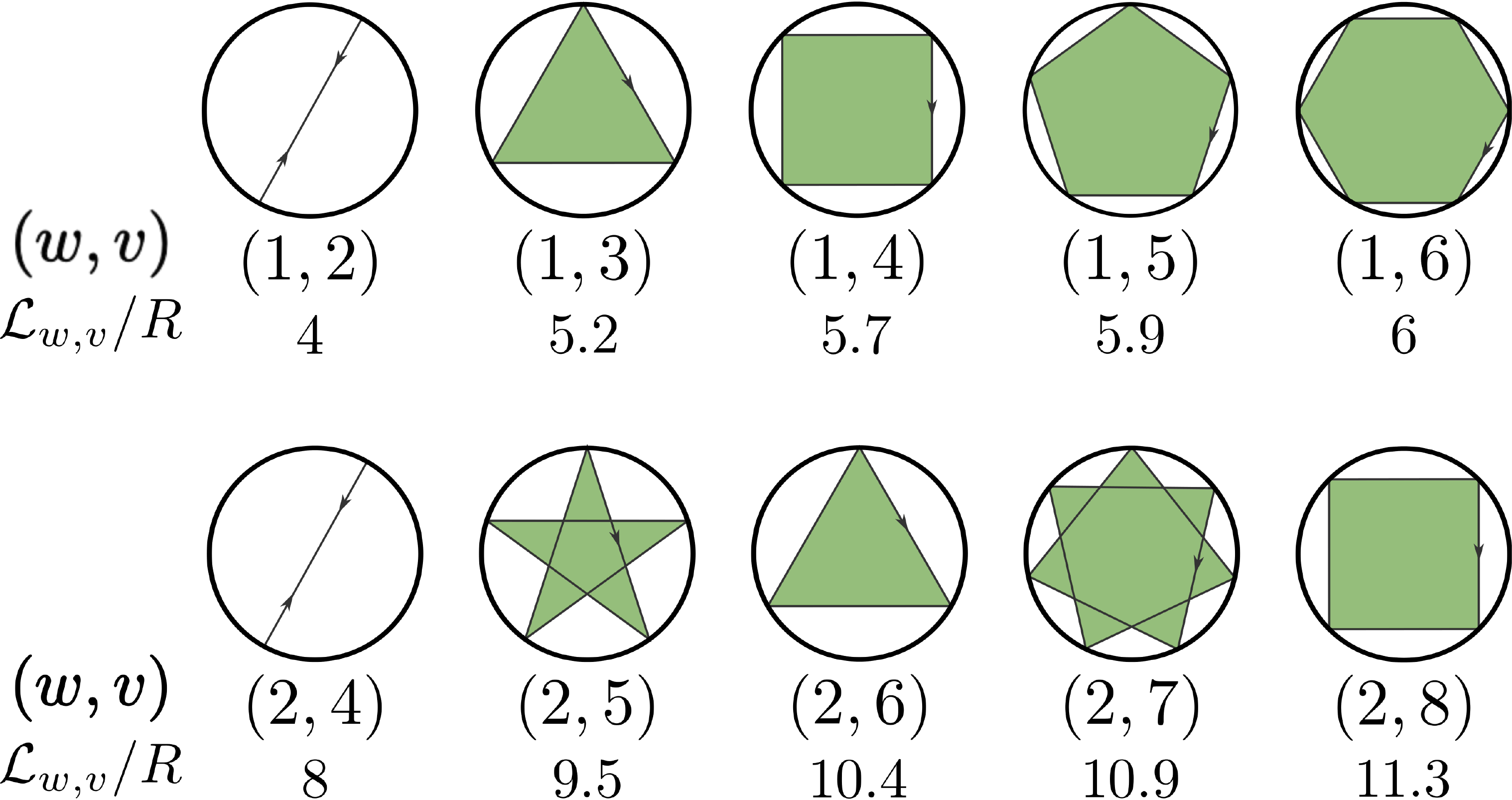}
\caption{Trajectories representing families of classical periodic orbits in the disk billiard.  
	  $w$ denotes the winding number, whereas $v$ labels the total number of boundary reflections.  
	  Enclosed areas are marked in green.
}
\label{fig:disk_orbits}
\end{figure}

The semiclassical properties of the disk cavity with infinite-mass-type edges have already 
been considered (for $B=0$) in \mbox{Refs.~[\onlinecite{epub12143}, \onlinecite{PhysRevB.84.075468}]}.  
In order to compute its magnetic properties within semiclassical approximation, we combine
these results with results adapted from \mbox{Ref.~[\onlinecite{Richter}]}, 
where $\chi_{\mathrm{sc}}^{\mathrm{osc}}$ for the Schr\"odinger disk billiard was derived.  
For the disk geometry, one can characterize the periodic-orbit families by 
their winding number $w$ and their total number $v$ of reflections at the boundary (with $v \geq 2 w$).  
The sign of $w$ defines the direction of rotation.  
A few representative periodic-orbit families are depicted in 
\mbox{Fig.~\ref{fig:disk_orbits}} for $w = 1, 2$, together with their lengths $\mathcal{L}_{w, v}$ 
and the enclosed areas $\mathcal{A}_{w, v}$ (green shaded).  
They can be calculated within basic geometry yielding\cite{Richter}
\begin{align}
 \label{eq:sc_d2}
  \mathcal{L}_{w, v} &= 2 v R \sin\left(\left|\pi \frac{w}{v}\right|\right),\\
\label{eq:sc_d3}
 \mathcal{A}_{w, v} &= \mathcal{A} \frac{v}{2 \pi} \sin\left(2\pi \frac{w}{v}\right),
\end{align}
with area $\mathcal{A} = \pi R^{2}$.
The trace over the pseudospin propagator for an orbit family characterized by the tupel $(w, v)$ can be calculated from 
\mbox{Eq.~(\ref{eq:sc_66})} and reads\cite{PhysRevB.84.075468,  epub12143}
\begin{align}
 \label{eq:sc_d4}
\mathrm{Tr}K_{w, v} = g\cos\left(v\, \theta_{w, v}\right)
	\begin{cases} 
		(-1)^{v/2} & \mathrm{for\ even\ }v,\\
		  0        & \mathrm{for\ odd\ } v 
	\end{cases},
\end{align}
with the reflection angle \mbox{$\theta_{w, v} = [\mathrm{sgn}(w)/2 - w/v]\pi$}.  
Due to pseudospin interference only orbits with an odd number of reflections contribute to the DOS,
in contrast to the corresponding Schr\"odinger system\cite{brack_sc, Richter}.
Therefore, the entire field-dependent, oscillatory contribution to the DOS reads
\begin{align}
 \label{eq:sc_d5}
\begin{split}
\rho_{\mathrm{sc}}^{\mathrm{osc}}(E, B) =& \frac{2}{\hbar v_{F}}\sqrt{\frac{k}{2 \pi}} \sum\limits_{w = 1}^{\infty}\sum\limits_{\substack{v \geq 2 w \\ \mathrm{even}}}^{\infty}
	  (-1)^{w + v/2} \frac{f_{w, v}}{v^{2}} \mathcal{L}_{w, v}^{3/2}\\
	&\times \sin\left(k\mathcal{L}_{w, v} + \frac{3}{4}\pi\right)  \mathcal{C}_{w, v}(B).
\end{split}
\end{align}
Owing to the rotational symmetry, the $B$-field induced modulation of each contribution is only due to dephasing 
between time-reversed orbits such that the magnetic phase factor reads\cite{Richter}
\begin{align}
 \label{eq:sc_d6}
 \mathcal{C}_{w, v}(B) = \frac{1}{2 \pi}\int\limits_{0}^{2 \pi}\mathrm{d}\varphi \cos\left({\frac{\mathcal{A}_{w, v}}{l_{B}^{2}}}\right) = \cos\left({\frac{\mathcal{A}_{w, v}}{l_{B}^{2}}}\right).
\end{align}
Together with \mbox{Eq.~(\ref{eq:sc_76})}, one then finds for the semiclassical approximation of the oscillatory contribution to the orbital magnetic susceptibility (in terms of $\chi_{0}$, \mbox{Eq.~(\ref{eq:m1_24})}):
\begin{align}
 \label{eq:sc_d7}
 \begin{split}
 \chi^{\mathrm{osc}}_{\mathrm{sc}}(\mu, B) =&-\chi_{0}(B)\times \frac{8 \pi^{3/2}}{3\zeta\left(3/2\right)} \frac{R}{l_{B}} \sqrt{k_{F}R}\\
	&\hspace{-1.5cm}\times \sum\limits_{w = 1}^{\infty}\sum\limits_{\substack{v \geq 2 w \\ \mathrm{even}}}^{\infty} \frac{(-1)^{w + v/2}}{v^{2}}\left(\frac{\mathcal{A}_{w,v}}{R^{2}}\right)^{2}\sqrt{\frac{R}{\mathcal{L}_{w,v}}} \\
	&\hspace{-1.5cm}\times \sin\left(k_{F}\mathcal{L}_{w, v} + \frac{3}{4}\pi\right)  \mathcal{C}_{w, v}(B) \mathrm{R}_{T}\left(\frac{L_{w,v}}{\hbar v_{F}}\right).
\end{split}
\end{align}

\begin{figure}[htbp]
 \centering
\includegraphics[width = 0.45\textwidth]{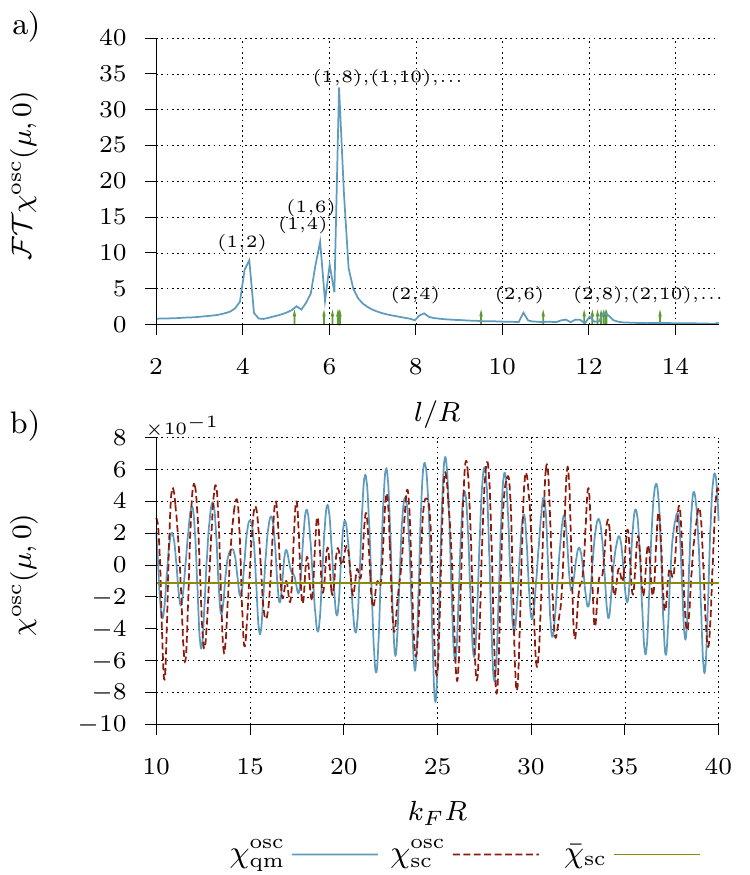}
\caption{
a) Length spectrum of $\chi^{\mathrm{osc}}$ calculated quantum mechanically from 
 the eigenenergies of a graphene disk, \mbox{Eq.~(\ref{eq:sc_d1})},
        at $\phi=0$, $1/\beta \approx10^{-3}\,t$ and using  $R \approx 200\,a$. 
     Peak positions correspond to orbit families $(w,v)$.
 Due to pseudospin interference, orbits with an odd number of reflections do not contribute.
Their lengths are marked by green arrows.  	
	b) Comparison of the semiclassical prediction (\ref{eq:sc_d7}) (red dashed line)
      for the orbital magnetic susceptibility with the quantum mechanical result (solid blue)
	at $B=0$ and $1/\beta \approx 10^{-3}\,t$.  
	The (green) horizontal line represents the typical value, \mbox{Eq.~(\ref{eq:sc_d10})},
        of the magnetic susceptibility.
	All susceptibilities are normalized by \mbox{$X = \chi_{0}(\phi_{0}/(2\mathcal{A}))$} 
         and $\sqrt{k_{F} R}$.  
}
\label{fig:disk}
\end{figure}

Since bouncing-ball orbits $(w,2w)$ do not enclose a finite area in the weak field limit they are
not considered in $\chi^{\mathrm{osc}}_{\mathrm{sc}}$, and we absorbed the factor $f_{w, v}=2$ 
into the overall prefactor. Expression (\ref{eq:sc_d7}) demonstrates that the confinement-induced 
magnetic response of an integrable geometry is parametrically larger (by a factor $\sqrt{k_{\rm F}R}$)
than the bulk value $\chi_0$.

Panel a) of \mbox{Fig.~\ref{fig:disk}} shows the length spectrum resulting from the Fourier transform of
the quantum-mechanical result for $\chi^{\mathrm{osc}}(\mu)$ at $B=0$.
One can clearly identify the peak positions with the lengths $\mathcal{L}_{w, v}$ of the shortest
contributing orbits as expected from the semiclassical formula (\ref{eq:sc_d7}).
Green arrows mark the lengths of those orbits that do not contribute due to destructive pseudospin
interference according to \mbox{Eq.~(\ref{eq:sc_d4})}.
Apparently, as visible in \mbox{Fig.~\ref{fig:disk}a)}, also bouncing-ball orbits 
\mbox{$(w, 2w)$} yield a contribution to the quantum mechanical result $\chi^{\mathrm{osc}}$, 
even though, according to Eq.~(\ref{eq:sc_d6}), their semiclassical contribution vanishes at weak fields
if bending of the trajectories is not included.
The temperature used in \mbox{Fig.~\ref{fig:disk}} is equivalent to a short
cut-off length of $\mathcal{L}_{c}\approx 1.5\,R$, implying that only the lowest harmonics contribute 
significantly to $\chi^{\mathrm{osc}}_{\mathrm{sc}}$. This may explain why the peak from the shortest
orbits, the bouncing-ball orbits is comparable to the other peaks.
The influence of this first peak causes small deviations between the semiclassical and the quantum mechanical 
result, as visible in \mbox{Fig.~\ref{fig:disk}b)}.  
There,  $\chi^{\mathrm{osc}}$ is 
normalized by $\sqrt{k_{F}R}$ and
\begin{align}
\label{eq:sc_X}
 X = (0.5\,\phi_{0}/\mathcal{A}) \chi_0 \approx-7.8\,R\cdot 10^{-5} \, .
\end{align}
Due to the divergent character of $\chi_{0}$ for small values of $\phi$
(see \mbox{Eq.~(\ref{eq:m1_24})}), the amplitude of the oscillations in $\chi^{\mathrm{osc}}$ appears 
to be smaller than the contribution from the filled valence band. 
Anyhow, one would not expect the quantum-mechanical and the semiclassical result to lie in perfect agreement with each other since the susceptibility as a second derivative is very sensitive to small deviations already 
on the level of the DOS.  
The length spectrum \ref{fig:disk}a) shows an accumulation of contributing orbits in the vicinity of $2\pi\,w$.  
This clustering of orbit families can be identified with the so called 'whispering gallery' modes, which yield a coherent contribution to $\rho^{\mathrm{osc}}_{\mathrm{sc}}$.  
Since these orbit families enclose nearly the whole disk area, 
i.e. $\mathcal{A}_{w, v} \approx \mathcal{A}\,w$, their contribution to $\chi^{\mathrm{osc}}_{\mathrm{sc}}$ 
converges as $(-1)^{v/2}/v^{2}$ for a fixed value of $w$ leading to an overall convergence of \mbox{Eq.~(\ref{eq:sc_d6})} at finite temperatures\cite{Richter}.  

As has been done for corresponding systems of parabolic dispersion\cite{Richter} 
one can calculate the typical value of the oscillatory susceptibility contribution defined by the 
root mean square of $\chi^{\mathrm{osc}}_{\mathrm{sc}}$ with respect to energy\cite{Richter}:
\begin{align}
 \label{eq:sc_d8}
\bar{\chi}_{\mathrm{sc}}(\mu, B) = \sqrt{\langle\left[\chi^{\mathrm{osc}}_{\mathrm{sc}}(\mu, B)\right]^{2}\rangle},
\end{align}
with
\begin{align}
 \label{eq:sc_d9}
 \hspace{-0.3cm}\langle\left[\chi^{\mathrm{osc}}_{\mathrm{sc}}(\mu, B)\right]^{2}\rangle 
      =\frac{1}{\Delta k_{F} R}\hspace{-0.5cm}\int\limits_{k_{F}R}^{k_{F}R + \Delta k_{F} R}\hspace{-0.5cm}\mathrm{d}k'_{F}R \left[\chi^{\mathrm{osc}}_{\mathrm{sc}}(E'_{F}, B)\right]^{2}.
 \hspace{-0.3cm}
\end{align}
The energy interval \mbox{$[k_{F} R, k_{F}R + \Delta k_{F}R]$} is chosen classically negligible but 
quantum mechanically large, i.e.\ $k_{F} R \gg \Delta k_{F}R \gg 2 \pi$.  
As a consequence, semiclassical off-diagonal terms \mbox{$\propto \sin(k_{F}\mathcal{L}_{w, v}) \sin(k_{F} \mathcal{L}_{w', v'})$}, where \mbox{$(w, v) \neq (w', v')$}, vanish under integration in \mbox{Eq.~(\ref{eq:sc_d9})}, 
whereas the diagonal terms yield a contribution of $1/2$.  
For a detailed discussion  see \mbox{Ref.~[\onlinecite{Richter}]}.  
In the zero field limit \mbox{Eq.~(\ref{eq:sc_d8})} simplifies to 
\begin{align}
 \label{eq:sc_d10}
\begin{split}
 \hspace{-0.4cm}\bar{\chi}_{\mathrm{sc}}(\mu, 0) =& -X\sqrt{k_{F}R}\times\frac{8\sqrt{\pi}}{3\zeta\left(\frac{3}{2}\right)} \\
      &\times \left[\frac{1}{2} \hspace{-0.4cm}\sum\limits_{\substack{w \\ v > 2 w, \mathrm{even}}}\hspace{-0.4cm} 
	  \frac{\mathrm{R}_{T}^{2}\left(\frac{\mathcal{L}_{w, v}}{\hbar v_{F}}\right)}{v^{4}}\frac{\left(\mathcal{A}_{w, v}/R^{2}\right)^{4}}{\mathcal{L}_{w, v}/R} 
	     \right]^{1/2}
\end{split} \hspace{-0.4cm}
\end{align}
in terms of $X$, Eq.~(\ref{eq:sc_X}).
Choosing a similar cut-off length as in \mbox{Fig.~\ref{fig:disk}}, i.e.\ \mbox{$\mathcal{L}_{c} = 1.5\,R$},
yields $\bar{\chi}_{\mathrm{sc}}(\mu, 0) \approx $ $ -0.11\,X\,\sqrt{k_{F}R}$, marked as a horizontal
line in Fig.~\ref{fig:disk}(b).
In contrast to that, a calculation\cite{Richter} yields for a circular quantum dot with parabolic dispersion 
 \mbox{$\bar{\chi}_{\mathrm{sc, 2DEG}} \approx 0.87\,\chi_{L}\,\left(k_{F} R\right)^{3/2}$}, 
where the Landau susceptibility $\chi_{L}$, \mbox{Eq.~(\ref{eq:m1_25})}, corresponds to $\chi_{0}$ in graphene.   

We additionally considered ring-shaped graphene billiards of various thickness with infinte-mass-type edges.  
As shown in \mbox{Ref.~[\onlinecite{PhysRevB.76.235404}]} this geometry can be quantized in Dirac approximation 
for arbitrary magnetic field strength yielding a condition similar to \mbox{Eq.~(\ref{eq:sc_d1})}.  
The comparison of $\chi_{\mathrm{qm}}^{\mathrm{osc}}$ with $\chi_{\mathrm{sc}}^{\mathrm{osc}}$ does not yield convincing coincidence in that case due to 
additional diffraction effects at the inner disk. These effects are beyond the leading-order 
semiclassical expansion considered in this work.  


\subsection{\label{sec:Main_2_rect}Rectangular billiard with zigzag and armchair edges}

\begin{figure}[htbp]
 \centering
\includegraphics[width = 0.3\textwidth]{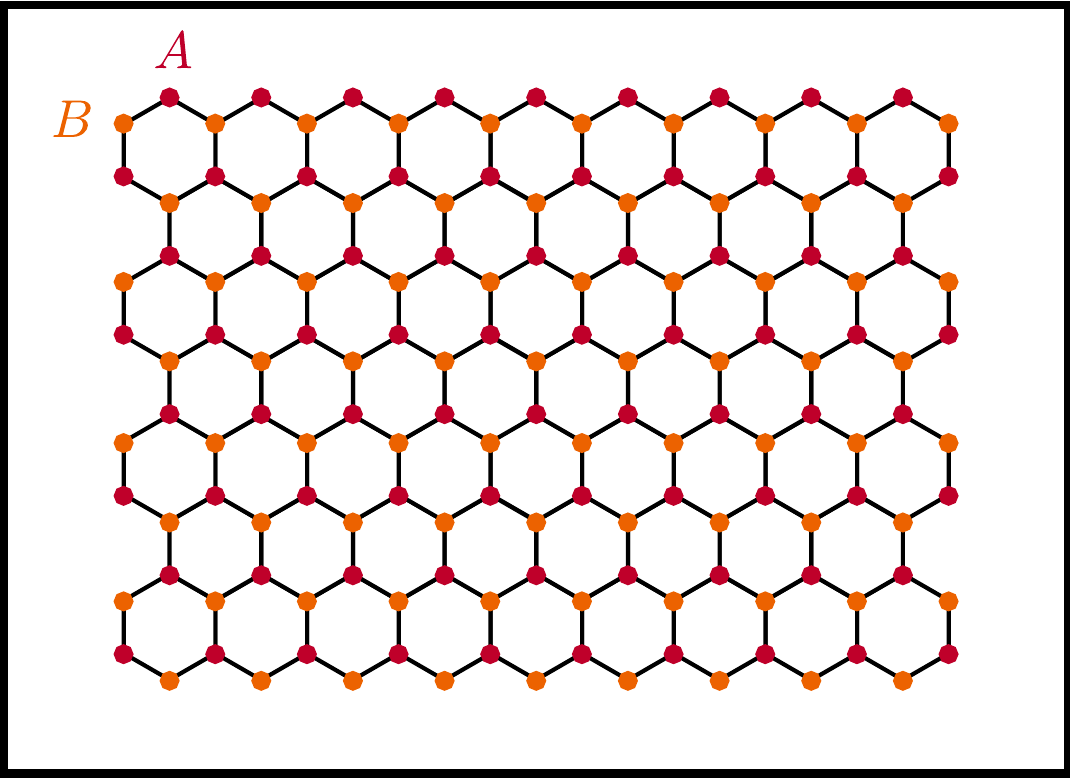}
\caption{Example of a typical rectangular graphene quantum dot with $\mathcal{L}_{\mathrm{ac}} = 11/\sqrt{3}\,a$ in $x$- and $\mathcal{L}_{\mathrm{zz}} = 9\,a$ in $y$-direction.  
}
\label{fig:rect_ex}
\end{figure}

The second fundamental system we consider is a rectangular graphene quantum dot with zigzag edges in 
$x$- and armchair edges in $y$-direction similar as shown in \mbox{Fig.~\ref{fig:rect_ex}}.  
The side lengths are labeled as $\mathcal{L}_{\mathrm{zz}}$ and $\mathcal{L}_{\mathrm{ac}}$, respectively, 
such that  \mbox{$\mathcal{A} = \mathcal{L}_{\mathrm{ac}}\mathcal{L}_{\mathrm{zz}}$}.
Similar to the comparable Schr\"odinger system, the Dirac equation for a rectangular graphene quantum dot 
cannot be solved analytically in the presence of a magnetic field.  
For this reason, we will calculate the eigenenergies numerically within tight-binding approximation to 
check the quality of the semiclassical prediction.  

\begin{figure}[htbp]
 \centering
\includegraphics[width = 0.45\textwidth]{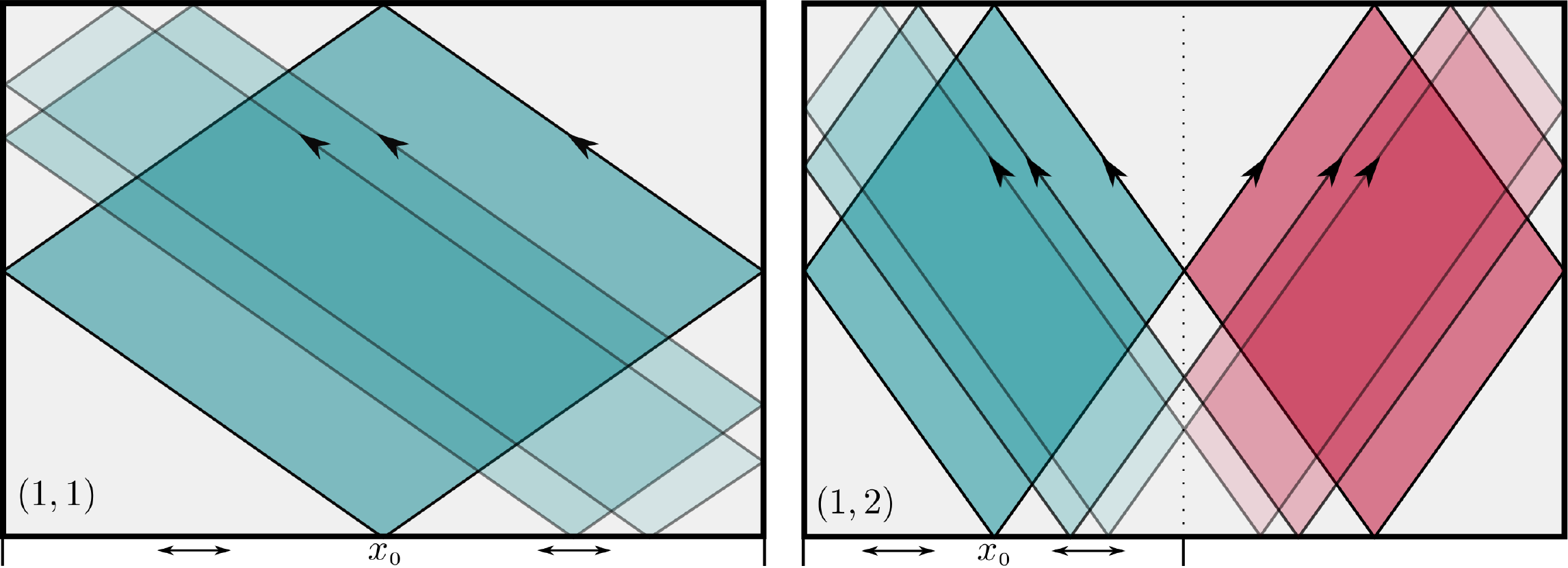}
\caption{Three representative members of the orbit families $(1,1)$ and $(1,2)$.  
The enclosed area varies depending on the position of the reflection point \mbox{$x_{0}\in\left[0,\mathcal{L}_{\mathrm{zz}}/n\right]$} at the lower boundary.  
In the case of the $(1,2)$ family the net directed enclosed area is zero due to opposite propagation 
direction along both trajectory parts.  
}
\label{fig:rect_generisch}
\end{figure}

From \mbox{Fig.~\ref{fig:rect_ex}} it is clear that opposite zigzag edges are built from different sublattices and lead to an additional sign of the reflection angle at one of the zigzag edges, as mentioned in 
\mbox{Sec. \ref{sec:Main_2}}.
The classical periodic paths in this system can be classified by the tuple \mbox{$(M, N) = (rm, rn)$} 
of their primitive reflection numbers $m$ and $n$ on the edges and their number of repetitions $r$.  
The corresponding orbit has $M$ bounces at the armchair and $N$ bounces at the zigzag edges in 
total and closes after $r$ repetitions.  \mbox{Figure \ref{fig:rect_generisch}} shows as 
examples members of the family $(1, 1)$ and $(1, 2)$, respectively.  
The members of one orbit family can be transformed into each other via translation of the reflection point \mbox{$x_{0} \in \left[0, \mathcal{L}_{\mathrm{zz}}/n\right]$} at the $x$-axis.  
Thus, all members of one family have the same path length\cite{brack_sc, Richter}
\begin{align}
 \label{eq:sc_re1}
 \mathcal{L}_{M, N} = 2 r \sqrt{\left(m \mathcal{L}_{\mathrm{zz}}\right)^{2} + \left(n \mathcal{L}_{\mathrm{ac}}\right)^{2}},
\end{align}
and only the enclosed area depends on the translational group element $x_{0}$.  
From \mbox{Refs.~[\onlinecite{brack_sc, Richter}]} follows
\begin{align}
\label{eq:sc_re2}
  \hspace{-0.35cm}\mathcal{A}_{M, N}(x_{0}) = \begin{cases}\frac{2 r}{m} \mathcal{L}_{\mathrm{ac}} x_{0}\left(1 - \frac{x_{0}}{\mathcal{L}_{\mathrm{zz}}} n\right) & \mathrm{if\ }m\cdot n\mathrm{\ odd,}\\
				0 &  \mathrm{if\ }m\cdot n\mathrm{\ even}.
                              \end{cases}
\end{align} 
As visible in \mbox{Fig.~\ref{fig:rect_generisch}}, the directed area does not vanish if
$m\cdot n$ is odd. The flux-dependent dephasing factor reads
\begin{align}
 \label{eq:sc_re3}
\mathcal{C}_{M, N}(B) =& \frac{n}{\mathcal{L}_{\mathrm{zz}}} \int\limits_{0}^{\mathcal{L}_{\mathrm{zz}}/n}\mathrm{d}x_{0}\,\cos\left(\frac{\mathcal{A}_{M, N}(x_{0})}{l_{B}^{2}}\right)\\
\begin{split}
	    =& \frac{\sqrt{\pi/2}}{\sqrt{\phi_{M, N}}}\left[\cos\left(\phi_{M, N}\right)\mathrm{C}\left(\sqrt{\phi_{M, N}}\right)\right.\\
	    &\hspace{0.5cm}\left. \,\, + \sin\left(\phi_{M, N}\right)\mathrm{S}\left(\sqrt{\phi_{M, N}}\right)\right].
\end{split}
\end{align}
The cosine Fresnel integral\cite{gradshtein} 
\mbox{$\mathrm{C}(x) = \sqrt{2/\pi}\int_{0}^{x}\mathrm{d}t\,\cos(t^{2})$} is defined analogous to 
$\mathrm{S}(x)$ in \mbox{Sec. \ref{sec:Main_1_cond}}. 
The phase
\begin{align}
 \label{eq:sc_re4}
 \phi_{M, N} =  \frac{\mathcal{A}_{M, N}(\mathcal{L}_{\mathrm{zz}}/(2 n))}{l_{B}^{2}} = \pi \frac{r}{m n} \varphi
\end{align}
corresponds to $2\pi$ times the flux through the area \mbox{$\mathcal{A}_{M, N}(\mathcal{L}_{\mathrm{zz}}/(2n)) = \mathcal{A} r/(2 mn)$}, 
which is enclosed by the time-reversed orbit partner with bounces at  \mbox{$x_{0} = \mathcal{L}_{\mathrm{zz}}/(2 n)$}.  
It can be directly proven that the enclosed area of these two orbits of the $(M,N)$ orbit family is maximum  
and therefore the action  (\mbox{Eq.~(\ref{eq:sc_70})}) stays extremal only for these two paths.
Corresponding to the Poincar\'{e}-Birkhoff theorem, these are the only members of the orbit family, which remain periodic in the presence of the perpendicular magnetic field\cite{Richter}.  
In contrast to rotational symmetric systems, the magnetic field factor is not only governed by the dephasing between time-reversed orbit twins but also 
due to dephasing of family members propagating in the same direction.  

The trace over the pseudospin-propagator is
 [\onlinecite{epub12143}, \onlinecite{PhysRevB.84.205421}] 
\begin{align}
 \label{eq:sc_re5}
\begin{split}
 \mathrm{Tr}K_{M, N} =& g(-1)^{r n}\cos\left(2 K \mathcal{L}_{\mathrm{zz}} r m- 2 r n \left|\theta_{\mathrm{zz}}\right|\right),
\end{split}
\end{align}
where $K \!=\! 4 \pi/(3 a)$ is the distance between the $\Gamma$- and one of the $K$-points in the first Brillouin zone.  
The reflection angle 
$ |\theta_{\mathrm{zz}}| \!=\! \arctan(M \mathcal{L}_{\mathrm{zz}} / (N\mathcal{L}_{\mathrm{ac}})) $
appears in \mbox{Eq.~(\ref{eq:sc_re5})} because the opposing zigzag edges are built from different 
sublattices (Fig.~\ref{fig:rect_ex}).  
On a microscopic scale the distance between both armchair edges can only take values \mbox{$\mathcal{L}_{\mathrm{zz}} = q\cdot a/2$}, \mbox{$q \in \mathbb{N}$}, yielding\cite{epub12143, PhysRevB.84.205421}
\begin{align}
 \label{eq:sc_re7}
  K\mathcal{L}_{\mathrm{zz}} = \begin{cases} 0 \mathrm{\,mod\,}2\pi & \mathrm{if\ }q\mathrm{\,mod\,}3 = 0,\\
				      \pi/3 \mathrm{\,mod\,}2\pi & \mathrm{otherwise}.
                               \end{cases}
\end{align}


Whenever the length of the zigzag edge is such that $q$ is a multiple of $3$, orbit families with \mbox{$(N \mathcal{L}_{\mathrm{ac}}/\mathcal{L}_{\mathrm{zz}}, N)$}, where $N$ is odd, are suppressed 
by the trace of the pseudospin-propagator and hence do not contribute to the DOS\cite{epub12143, PhysRevB.84.205421}.  
Furthermore, when \mbox{$\mathcal{L}_{\mathrm{ac}} = \mathcal{L}_{\mathrm{zz}}$} on a macroscopic scale, 
bouncing ball orbits with $(0, M)$ and $(N, 0)$ cancel each other exactly if $M$ and $N$ are 
odd\cite{epub12143, PhysRevB.84.205421}, respectively.  
Combining these considerations with the results for a rectangular Schr\"odinger system\cite{Richter, brack_sc} we 
find the field-dependent expression 
\begin{align}
 \label{eq:sc_re8}
\begin{split}
 \hspace{-0.3cm}\rho^{\mathrm{osc}}_{\mathrm{sc}}(E, B) =& \frac{\mathcal{A}}{\hbar v_{F}} \sqrt{\frac{k}{2\pi^{3}}}
		\sum\limits_{r = 1}^{\infty}\sum\limits_{\substack{m, n = 0 \\ m = 0 \vee n = 0}}^{\infty}\frac{f_{n, m}}{\sqrt{\mathcal{L}_{M, N}}}\\
	    &\times\cos\left(k\mathcal{L}_{M, N} - \frac{\pi}{4}\right)\mathrm{Tr}K_{M, N}\mathcal{C}_{M, N}(B). \hspace{-0.33cm}
\end{split}
\end{align}
The factor $f_{M,N}=2$ whenever there exists a time-reversed version of the orbit family $(M, N)$ and \mbox{$f_{M,N} = 1$} for bouncing-ball orbits.  
In order to calculate $\chi^{\mathrm{osc}}_{\mathrm{sc}}$ we take the second derivative of the field
factor,
\begin{align}
 \label{eq:sc_re9}
 \mathcal{C}''_{M, N}(B) = -\left(\frac{2\pi \mathcal{A}_{M, N}}{\phi_{0}}\right)^{2} \frac{\sqrt{\pi/2}}{4} \times \tilde{\mathcal{C}}_{M,N}\left(\phi_{M, N}\right),
\end{align}
with $\phi_{M, N}$ in \mbox{Eq.~(\ref{eq:sc_re4})} and 
\begin{align}
 \label{eq:sc_re10}
\hspace{-0.25cm}\begin{split}
 \tilde{\mathcal{C}}_{M,N}\left(x\right) = & 
					   \sqrt{\frac{2}{\pi}}\frac{3}{x^{2}}
-\frac{\mathrm{C}(\sqrt{x})}{x^{5/2}}\left[\left(3 - 4 x^2\right)\cos(x) + 4 x \sin(x)\right]\\
					   & 
  \qquad \quad - \frac{\mathrm{S}(\sqrt{x})}{x^{5/2}}\left[\left(3 - 4 x^2\right)\sin(x) - 4 x \cos(x)\right]
\end{split}\hspace{-0.25cm}
\end{align}
In the zero field limit $\tilde{C}_{M, N}$ converges to the value \mbox{$(32/15)\sqrt{2/\pi}$}.  
For $\chi^{\mathrm{osc}}_{\mathrm{sc}}$ we find, according to \mbox{Eq.~(\ref{eq:sc_76})},
\begin{align}
 \label{eq:sc_re11}
\hspace{-1.5cm}\begin{split}
 \chi^{\mathrm{osc}}_{\mathrm{sc}}(\mu, B) =& -\chi_{0}(B)\times \frac{\sqrt{2}2\pi}{3\zeta(3/2)}\left(\frac{\mathcal{L}_{\mathrm{zz}}}{\mathcal{L}_{\mathrm{ac}}}\right)^{2} \frac{\mathcal{L}_{\mathrm{ac}}}{l_{B}} \sqrt{k_{F} \mathcal{L}_{\mathrm{ac}}}\\
	  &\hspace{-1.5cm}\times\sum\limits_{r = 1}^{\infty}\sum\limits_{\substack{m, n = 1 \\ m\cdot n \mathrm{\,odd}}}^{\infty}\hspace{-0.15cm}\frac{\mathrm{Tr}K_{M, N}}{g}\left(\frac{\mathcal{A}_{M, N}}{\mathcal{A}}\right)^{2}\sqrt{\frac{\mathcal{L}_{\mathrm{ac}}}{\mathcal{L}_{M, N}}}\\
	  &\hspace{-1.5cm}\times\cos\left(k_{F}\mathcal{L}_{M, N} - \frac{\pi}{4}\right)\! \mathrm{R}_{T}\left(\!\frac{\mathcal{L}_{M, N}}{\hbar v_{F}}\!\right)\! \tilde{\mathcal{C}}_{M, N}\left(\!\phi_{M, N}\!\right).
\end{split}\hspace{-1.5cm}
\end{align}
The factor $f_{M,N}=2$ for all contributing orbit families, is absorbed in the prefactor.  
The squared aspect ratio  ${\cal L}_{\mathrm{zz}}/{\cal L}_{\mathrm{ac}}$ enters the prefactor yielding a strong dependence of the susceptibility on the geometry of the system.  

\begin{figure}[htbp]
 \centering
\includegraphics[width = 0.45\textwidth]{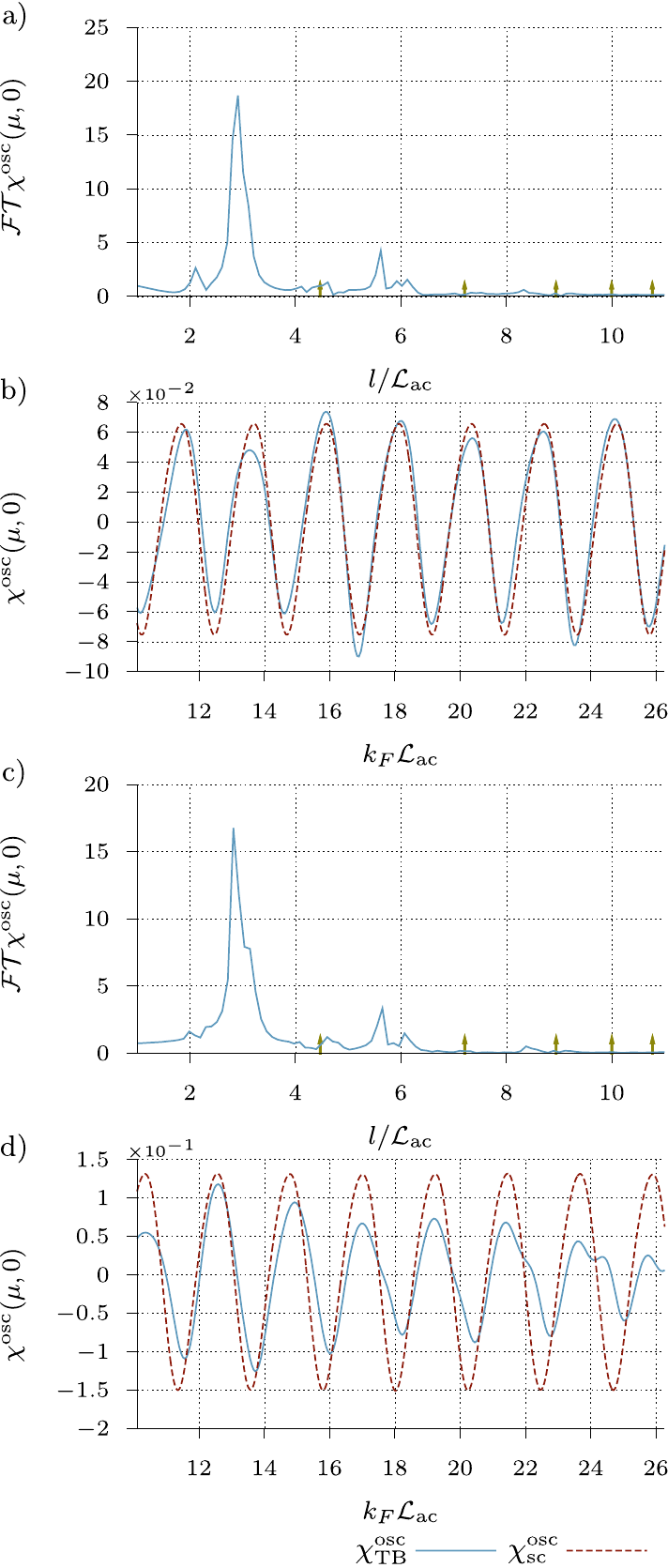}
\caption{Panels a) and b): Length spectra calculated from the Fourier transform of the tight-binding results
for $\chi^{\mathrm{osc}}$; panels b) and d): Comparison of $\chi^{\mathrm{osc}}$ in semiclasscial approximation 
with corresponding numerical tight-binding results.
Results shown in  a) and b) are obtained for a rectangular graphene quantum dot 
with \mbox{$K\mathcal{L}_{\mathrm{zz}} = 0\,\mathrm{mod}\,2\pi$} 
and  side lengths of \mbox{$\mathcal{L}_{\mathrm{zz}} = 201\,a$} and \mbox{$\mathcal{L}_{\mathrm{ac}} \approx 201.207\,a$}, respectively.  
Panels c) and d) display results for a cavity with side lengths  
\mbox{$\mathcal{L}_{\mathrm{zz}} = 202\,a$} and \mbox{$\mathcal{L}_{\mathrm{ac}} \approx 202.073\,a$} such that \mbox{$K\mathcal{L}_{\mathrm{zz}} = \pi/3\,\mathrm{mod}\,2\pi$}.  
Green arrows in panels a) and c) mark positions of orbits that do not contribute to the susceptibility 
due to their vanishing directed enclosed area $\mathcal{A}_{M, N}$.  
$\chi^{\mathrm{osc}}$ is normalized by $X/\sqrt{k_{E}\mathcal{L}_{\mathrm{ac}}}$ in all cases, with $X$ as
defined in \mbox{Eq.~(\ref{eq:sc_X})}.
}
\label{fig:rectangles}
\end{figure}

Panels a) and c) of \mbox{Fig.~\ref{fig:rectangles}} show the length spectra obtained after Fourier
transform from $\chi^{\mathrm{osc}}$ calculated from the tight-binding eigenenergies of two
rectangular graphene quantum dots with 
\mbox{$\mathcal{L}_{\mathrm{ac}} = 201.207\,a$}, \mbox{$\mathcal{L}_{\mathrm{zz}} =  201\,a$} and  \mbox{$\mathcal{L}_{\mathrm{ac}} = 202.073\,a$}, \mbox{$\mathcal{L}_{\mathrm{zz}} = 202\,a$}, respectively.  
The green arrows mark the position of orbit families which are semiclassically predicted not to 
contribute to $\chi^{\mathrm{osc}}$.
These length spectra are not as smooth as the one obtained for the graphene disk with infinite mass
boundaries, \mbox{Fig.~\ref{fig:disk}a)}, since
the region of linear dispersion cannot be extended arbitrarily in tight-binding approximation.  
Still, one can clearly identify the peaks in \mbox{Fig.~\ref{fig:rectangles}a)} and c) with the lengths of
contributing orbits, such that the comparison of $\chi^{\mathrm{osc}}_{\mathrm{TB}}$, with
$\chi^{\mathrm{osc}}_{\mathrm{sc}}$, \mbox{Eq.~(\ref{eq:sc_re11})},
in \mbox{Fig.~\ref{fig:rectangles}b)} and d) shows convincing agreement.  
In these cases the thermal energy is \mbox{$1/\beta = 10^{-3}\,t$} corresponding to the cut-off length 
\mbox{$\mathcal{L}_{c}\approx 1.5\,\mathcal{L}_{\mathrm{ac}}$}.
The normalization factor $X/\sqrt{k_{F}\mathcal{L}_{\mathrm{ac}}}$ is the same as the one chosen in
\mbox{Subsec.~\ref{sec:Main_2_disk}}, with $X$ defined by \mbox{Eq.~(\ref{eq:sc_X})} 
and $\mathcal{L}_{\mathrm{ac}} \approx R$, i.e. all parameters are similar to the disk.    
The amplitudes of the oscillation between para- and diamagnetic behavior of $\chi^{\mathrm{osc}}$ in the case of the rectangular quantum dots [\mbox{Fig.~\ref{fig:rectangles}b)} and d)] are similar to the amplitude of  $\chi^{\mathrm{osc}}$ 
for the circular quantum dot [\mbox{Fig.~\ref{fig:disk}b)}].  
Though the agreement of the oscillation frequencies of $\chi^{\mathrm{osc}}_{\mathrm{TB}}$ and $\chi^{\mathrm{osc}}_{\mathrm{sc}}$ in \mbox{Fig.~\ref{fig:rectangles}b)} and d) are convincing 
the tight-binding result in panel d) exhibits an additional modulation of the oscillations for \mbox{$k_{F}\mathcal{L}_{\mathrm{ac}} > 22$} which 
are not contained in the semiclassical approximation. 
In the corresponding energy range, the Dirac model and therefore the semiclassical approximation reaches 
the limit of validity\cite{epub12143} in describing the energy spectrum of graphene when $k\,a\lesssim 1$. 
Though the lengths $\mathcal{L}_{\mathrm{ac}, \mathrm{zz}}$ of the system considered in
\mbox{Fig.~\ref{fig:rectangles}a), b)} ($K\,\mathcal{L}_{\mathrm{zz}} = 0\,\mathrm{mod}\,2\pi$) are
 only one row of atoms shorter on each side than the system 
considered in \mbox{Fig.~\ref{fig:rectangles}c), d)} ($K\,\mathcal{L}_{\mathrm{zz}} = \pi/3\,\mathrm{mod}\,2\pi$), the oscillation amplitude of $\chi^{\mathrm{osc}}$ differs by one order of magnitude.  
This is due to the suppression of orbit families with
$(N\,\mathcal{L}_{\mathrm{ac}}/\mathcal{L}_{\mathrm{zz}}, N)$, where $N$ is odd, as noted above.
Since the aspect ratio is not perfectly integer, those orbit families still yield a small contribution to
$\rho^{\mathrm{osc}}$, and correspondingly to $\chi^{\mathrm{osc}}$,
 and appear in the length spectrum, \mbox{Fig.~\ref{fig:rectangles}b)}.

\section{\label{sec:Conclusion} Summary and Outlook}

In this work we focused on the orbital magnetic properties of non-interacting ballistic bulk graphene  and
in particular on confined graphene-based systems with regular classical dynamics.  
To this end we considered the magnetic susceptibility $\chi$, calculated in the grand canonical ensemble,
in the energy region of linear dispersion.

In the first part of this paper, we considered bulk graphene. There the orbital magnetic response 
distinctly depends on the particular energy scales involved, namely the different energy regimes
associated with temperature, chemical potential and magnetic field that we considered with a 
comparitive look.

In a first step we derived the temperature-independent susceptibility contribution $\chi_0$ 
from the filled valence band, {\em i.e.} the graphene analogue to the Landau suceptibility $\chi_{L}$ of
an  ordinary two-dimensional electron gas.  
We found for $\chi_{0}$ the well-known diamagnetic $-B^{-1/2}$ behavior, 
assuming, in accordance with literature, the valence band to be linear and extended, 
so $\chi_{0}$ cannot be directly compared to realistic tight-binding calculations even for very large systems. 
Still, for finite temperatures we found the total orbital magnetic susceptibility 
$\chi = \chi_{0} + \chi_{T}$ to be regular in the limit $B \rightarrow 0$ for $T \neq 0$.
We compared our analytic results for the temperature-dependent part 
$\chi_{T}$ of the susceptibility 
with the results from literature and to numerical tight-binding calculations. The latter
were performed for finite nanostructures of mesoscopic dimensions, also in view of the confinement effects
later addressed. Still, the magnetic response of these graphene cavities also exhibits
bulk-like features, and we discussed initially the necessary conditions for comparision of our 
analytic bulk calculations with results for finite systems.  

In the presence of a finite magnetic field the features of the susceptibility depend on the 
relative size of the associated Landau level spacing $\Delta_{LL}$, the chemical potential $\mu$
and the thermal energy $k_{\rm B}T$. If the latter is the smallest scale, we distinguish two regimes
(see Fig.\ 1):

For $\mu > \Delta_{LL}$
we obtained the typical $\mu^{2}$- and $1/B$-equidistant, oscillatory behavior of $\chi_{T}$,
similar to de Haas-van Alphen oscillations in two-dimensional electron gases.  
Though the corresponding numerically calculated magnetic response for the finite systems 
exhibits, as a signature of the confinement, a richer oscillatory structure in this regime, 
there is a clear coincidence of clustered peaks with the pattern of $\chi_{T}$ in the bulk system.  
This becomes even more obvious by raising the temperature in the numerics, since then the finite-size
contributions are damped out and only Landau level signatures remain.  
The amplitudes of these  de Haas-van Alphen-type oscillations in graphene are one order of magnitude 
larger than the diamagnetic $\chi_{0}$ for the considered parameters, implying that the total orbital 
magnetic susceptibility oscillates between para- and diamagnetic behaviour as a function of 
$\mu$ and $B$, respectively.  

For  $\mu < \Delta_{LL}$ the term $\chi_{T}$, and therefore $\chi$, is an exponentially 
decaying function of the magnetic field and diamagnetic.  
For field values high enough such that bulk effects dominate over finite-size signatures,
the numerically calculated susceptibility of the quantum dots coincides very well with the analytic
results.  

If $k_{\rm B}T$ is larger than $\Delta_{LL}$,  
$\chi_{T}$ is a smooth function of temperature, chemical potential and magnetic field and 
shows paramagnetic behavior with values \mbox{$\lesssim 0.4\,|\chi_{0}|$}.  
Therefore, $\chi=\chi_0+\chi_{T}$ is diamagnetic and appears even to be independent of the magnetic 
field for arbitrary $\mu$. 
At the Dirac point ($\mu=0$), $\chi_T$, and correspondingly $\chi$, follow a Curie-type $T^{-1}$ power law which 
is confirmed by our numerical data for the (triangular) quantum dots at finite temperature.
For $1/\beta \lesssim t$, deviations between $\chi_{T}$ for the bulk and the finite systems 
appear due to the increasing relevance of finite-size signatures in this limit.  
We also analytically confirmed the well-known $\delta(\mu)$ singularity
\cite{PhysRev.104.666, PhysRevB.20.4889, PhysRevB.75.235333, PhysRevB.76.113301, PhysRevLett.102.177203, JPSJ.80.114705, 1751-8121-44-27-275001, PhysRevB.83.235409, PhysRevB.80.075418} 
of $\chi$ at zero temperature.

Through the confirmation of the analytic results for extended graphene with numerical data 
of finite quantum dots, we could analyze the importance of bulk effects in finite system on the one hand
and distinguish them from true confinement effects on the other hand.  
As one interesting aspect we found $\chi_{T}/\chi_{0}$ of the triangular quantum dot with zigzag edges 
to be smaller than  $\chi_{T}/\chi_{0}$ for the armchair quantum dot with same parameters.  
This is due to the zigzag edge state and the lower average energy in that case.
Moreover, especially in the energy range, where oscillations occur in $\chi_{T}$, 
the influence of the boundary is clearly observable.  

In the second, major part of this work we then analyzed in detail such confinement effects.
To this end we considered two representative geometries, a disk-shaped and a rectangular graphene cavity.
We derived a generic analytic expression for the oscillatory part $\chi^{\mathrm{osc}}$ of the orbital
magnetic susceptibility based on results [\onlinecite{Richter}] for the 
susceptiblity of confined electron gases and working out generalizations to finite $B$-fields of
semiclassical expressions for the field-free density of states for graphene cavities,
\mbox{Refs.~[\onlinecite{epub12143, PhysRevB.84.075468, PhysRevB.84.205421}]}.
We demomstrated that graphene specific edge effects depending on the type of the  boundaries
enter the semiclassical expressions, and thereby orbital magnetism, through phases associated
with the pseudospin propagator.   
This semiclassical approximation applies in particular to the low-field regime, where bulk contributions
are suppressed and the energy spectrum (and correspondingly  the orbital susceptibility) is governed 
by finite-size effects.  

We found good agreement of our semiclassical approach with the quantum mechanical results
for $\chi^{\mathrm{osc}}$ based on the calculation of the eigenergies for circular graphene quantum
dots with infinite-mass type edges.  
The Fourier transform of $\chi^{\mathrm{osc}}_{\mathrm{qm}}$ with respect to the energy
 yielded  a length spectrum with relatively sharp peaks reflecting the underlying classical orbit 
dynamics of this system. 
We showed that orbits with odd number of reflections are suppressed as it is predicted in our
semiclassical approach due to destructive pseudospin interferences.  
This is distinctly different from the corresponding case of the electron gas system\cite{Richter}, 
where all non self-retracing orbits yield a contribution to $\chi^{\mathrm{osc}}_{\mathrm{sc}}$.  
We found the typical value for  $|\chi^{\mathrm{osc}}|$ to scale like $\sqrt{k_{F} R}$.  
Hence, similar as in Ref.~[\onlinecite{Richter}] $|\chi^{\mathrm{osc}}|$ can be larger 
than $X = \chi_{0}(0.5\,\phi_{0}/\mathcal{A})$. In contrast,
the amplitudes of the $\chi^{\mathrm{osc}}$-oscillations show the same scaling behavior, but
appear to be of the same order of magnitude than $X$.
Similar agreement was found for rectangular-shaped graphene quantum dots, where we
compared the semiclassical predictions with numerical tight-binding calculations.
Depending on the length of the zigzag edges, the strength of the oscillatory modulations in 
$\chi$ were found to differ by one order of magnitude due to destructive pseudospin interferences. 

We studied the magnetic response for individual systems, including the typical susceptibility,
within the grand canonical formalism. To compute the average response of an ensemble of nanostructures,
a canonical treatment starting from the free energy 
instead of the grand potential is required~\cite{Imry-book}.
Along the lines of \cite{PhysRevLett.74.383,Richter}, and with the semiclassical expressions for graphene 
derived here, it appears straight forward to compute the ensemble-averaged susceptibility.

A further interesting aspect concerns the role of disorder for orbital magnetism in graphene,
both for the bulk and confined case. Again, previous work \cite{McCann,McCann2} for the 2d Schr\"odinger 
case, covering the entire disorder range from clean to diffusive, could act as a guideline.

Our overall analysis demostrates pronounced confinement effects on orbital magnetism in graphene-based 
nanosystems that dominate the bulk response in wide parameter regimes. However, our approach is
based on non-interacting models for graphene, as most of the works on orbital magnetism in graphene.
An exception is Ref.\ \onlinecite{PhysRevB.80.075418} where interaction effects are considered at 
$T=0$, however only to first order in the Coulomb repulsion.
The physics of conventional two-dimensional electron 
systems shows that, while non-interacting terms are also crucial there, contributions from electron-electron
interactions can usually not be disregarded. For instance, for the two-dimensional bulk Aslamazov and 
Larkin computed interaction corrections to the Landau susceptibility \cite{Aslamazov} (see also
Ref.~\cite{Ullmo1} for a semiclassical treatment).
Moreover, this work demonstrated that higher-order diagrams are essential for an
appropriate perturbative treatment of interaction effects, a treatment that is missing for graphene.
In Ref.~\cite{Ullmo2} it was furthermore shown that additional confinement-mediated 
interaction contributions to the susceptibility of 2d electron systems can be of the same order as 
those from the non-interacting model. To generalize such an analysis in terms
of interaction effects for graphene is beyond the scope of the present work.
Hence this interesting and challenging question is left for future research.

\section{Acknowledgments}

We thank Inanc Adagideli and J\"urgen Wurm for useful discussions and Viktor Kr\"uckl for help in 
numerical implementations.
This work was funded by the {\em Deutsche Forschungsgemeinschaft} through GRK 1570:
{\em Electronic Properties of Carbon Based Nanostructures}.

\appendix
\section{\label{sec:App_Int_Fresnel}Transformation of the Fresnel integral}

Starting with the definition of the Fresnel integral\cite{gradshtein}, \mbox{$\mathrm{S}(x) =
\sqrt{2/\pi}\int_{0}^{x}\mathrm{d}t\,\sin\left(t^{2}\right)$}, one 
finds after substituting \mbox{$t^{2} = \tau$} and using the relation\cite{gradshtein}
\begin{align}
 \label{eq:ap_1}
\frac{1}{z^{\alpha}} = \frac{1}{\Gamma\left(\alpha\right)}\int\limits_{0}^{\infty}\mathrm{d}u\,\frac{\ex^{-zu}}{u^{1-\alpha}},
	    \quad\mathrm{Re}\,z > 0, \mathrm{Re}\,\alpha > 0,
\end{align}
where \mbox{$z = \tau$} and \mbox{$\alpha = 1/2$},
\begin{align}
 \label{eq:ap_2}
\mathrm{S}\left(\left|x\right|\right) &= \frac{1}{\sqrt{2\pi}\Gamma\left(\frac{1}{2}\right)} \mathrm{Im}\left[
		\int\limits_{0}^{\infty}\mathrm{d}u\, \frac{1}{\sqrt{u}}
		\int\limits_{0}^{x^{2}}\mathrm{d}\tau\, \ex^{-(u-\im)\tau}\right]\\
\label{eq:ap_3}
      &= \frac{1}{\sqrt{2}\pi}\mathrm{Im}\left[
		 \int\limits_{0}^{\infty}\mathrm{d}u\, \frac{1}{\sqrt{u}(u-\im)}
		-\int\limits_{0}^{\infty}\mathrm{d}u\, \frac{\ex^{-(u-\im)x^{2}}}{\sqrt{u}(u-\im)}\right].
\end{align}
The first term in \mbox{Eq.~(\ref{eq:ap_3})} yields $1/2$ and represents the smooth part of the Fresnel integral.  
Using\cite{gradshtein}
\begin{align}
 \label{eq:ap_4}
 \mathrm{U}\left(1-\alpha; 1-\alpha; x\right) = \frac{x^{\alpha}}{\Gamma\left(1-\alpha\right)}
		  \int\limits_{0}^{\infty}\mathrm{d}t\,\frac{\ex^{-t} t^{-\alpha}}{t+x}
\end{align}
one finds (\mbox{$\alpha = 1/2$} and \mbox{$t = u x^{2}$}) \mbox{$\mathrm{S}(|x|) = 1/2 + \tilde{\mathrm{S}}(x)$} with 
\begin{align}
 \label{eq:ap_5}
\tilde{\mathrm{S}}\left(x\right) =  - \frac{1}{\sqrt{2\pi}}\mathrm{Im}
	    \left[\ex^{\im \frac{\pi}{4}}\ex^{\im x^{2}} \mathrm{U}\left(\frac{1}{2}; \frac{1}{2}; -\im x^{2}\right)\right].
\end{align}

\section{\label{sec:App_Int_Omega}Transformation of $\tilde{\Omega}_T$ for $\alpha, \gamma > 1$}
In order to calculate $\tilde{\Omega}_{T}$ as given in \mbox{Eq.~(\ref{eq:m1_49})} for $\alpha, \gamma > 1$ it is 
useful to apply the Taylor series representations of the logarithmic and exponential function yielding
\begin{align}
 \label{eq:ap_6}
\begin{split}
 \tilde{\Omega}_{T} - \hat{\Omega}_{T} =& g\frac{\varphi}{\beta}
	\sum_{s\pm 1}\sum\limits_{\substack{n=1 \\ m = 1}}^{\infty}\frac{(-1)^{m}}{m}\ex^{s\frac{\gamma}{\alpha}\,m}\\
	&\times\sum\limits_{k=0}^{\infty}
	 \frac{(-1)^{k}}{\Gamma\left(k+1\right)}( \sqrt{2}\gamma\, m)^{k}\,n^{\frac{k}{2}}.  
\end{split}
\end{align}
In the next step we interchange the order of summation\cite{PhysRevB.75.115123}, which can be done without causing correction terms in this particular situation\cite{elizalde2012ten}.  
Computing the sum over the Landau index $n$ first yields\cite{gradshtein} \mbox{$\sum_{n=1}^{\infty}n^{k/2}=\zeta(-k/2)$}, where $\zeta(z)$ is the Riemann zeta function.  
With use of\cite{gradshtein}
\begin{align}
 \label{eq:ap_7}
 \zeta(z) = \frac{1}{\Gamma(z)}\int_{0}^{\infty}\mathrm{d}t\,\frac{t^{z-1}}{\ex^{t}-1}
\end{align}
\mbox{Eq.~(\ref{eq:ap_6})} transforms to 
\begin{align}
\label{eq:ap_8}
\begin{split}
  \tilde{\Omega}_{T} - \hat{\Omega}_{T} =& \,g\frac{\varphi}{\beta}
	\sum_{s\pm 1}\sum\limits_{m=1}^{\infty}\frac{(-1)^{m}}{m}\ex^{s\frac{\gamma}{\alpha}\,m}\\ 
	& \hspace*{-0.45cm}\times\sum\limits_{k=0}^{\infty}\frac{(-1)^{k}}{\Gamma\left(k+1\right)\Gamma\left(-\frac{k}{2}\right)}
	\int_{0}^{\infty}\!\!\! \mathrm{d}t\,\frac{\left(\frac{\sqrt{2}\gamma\, m}{\sqrt{t}}\right)^{k}}{t\left(\ex^{t}-1\right)}.
\end{split}
\end{align}
We substitute \mbox{$t = 2\gamma^{2}\,m^{2}\cdot y = u\cdot y$} such that the integral in \mbox{Eq.~(\ref{eq:ap_8})} can be approximated by
\begin{align}
 \label{eq:ap_9}
  \int\limits_{0}^{\infty}\mathrm{d}y\,\frac{y^{-\frac{k}{2}-1}}{\exp\left(u\,y\right) - 1}
		\stackrel{\gamma > 1}{\approx}&  \int\limits_{0}^{\infty}\mathrm{d}y\,y^{-\frac{k}{2}-1}\ex^{-u\,y} = \Gamma\left(-\frac{k}{2}\right) u^{\frac{k}{2}}.
\end{align}
Calculating subsequently the sums over $k$ and $m$ in \mbox{Eq.~(\ref{eq:ap_8})} finally yields
\begin{align}
 \label{eq:ap_10}
 \tilde{\Omega}_{T} - \hat{\Omega}_{T} \approx-g\frac{\varphi}{\beta} \sum\limits_{s\pm 1} \ln\left[1 + \ex^{-\sqrt{2}\gamma + s\frac{\gamma}{\alpha}}\right].
\end{align}


\nocite{*}

\bibliography{mybib}

\end{document}